\documentclass{article}
\usepackage{arxiv}
\usepackage[toc,page]{appendix}
\usepackage{amsmath,amssymb}
\usepackage[english]{babel}
\usepackage[dvipsnames]{xcolor}  
\usepackage{array}
\usepackage{siunitx}       
\usepackage[mathscr]{euscript}
\usepackage{tensors}       
\usepackage{multirow}
\usepackage{float}
\usepackage{graphicx}
\usepackage[inkscapelatex=false]{svg}
\usepackage{nccmath}
\usepackage[export]{adjustbox}
\usepackage{setspace}
\usepackage{caption}
\usepackage{subcaption}
\usepackage{soul}
\usepackage{url}
\usepackage{tikz}
\usetikzlibrary{arrows.meta, positioning}
\usepackage{pgfplots}
\usepackage{todonotes}
\usepackage{siunitx}
\usepackage{soul}
\usepackage[ruled,vlined]{algorithm2e}
\usepackage{setspace}
\usepackage{enumitem}
\usepackage{threeparttable}

\title{Machine Learning-Accelerated Time Integration of Plasticity Models}

\author{
Nasrin Talebi$^{1,}$\thanks{Corresponding author: \texttt{nasrin.talebi@chalmers.se}}\,, \,
Magnus Ekh$^{1}$, \,
Knut Andreas Meyer$^{1}$
\\
$^{1}$Division of Computational Mechanics and Materials Engineering, Department of Mechanical \\
Engineering, Chalmers University of Technology, SE-412 96 Gothenburg, Sweden
}

\begin{document}
\maketitle

\begin{abstract}
Finite element simulations of structures with nonlinear material behavior require advanced material models to provide accurate predictions. However, the computational costs of these models can be high, as they solve coupled differential algebraic equations at each integration point, in each equilibrium iteration, in every time step. In this study, we propose a machine learning-based framework to accelerate these computations by explicitly calculating the state variable updates with neural networks, enabling large time steps with low computational costs. The neural networks operate on invariants, and the necessary and sufficient evolution directions are determined analytically based on the training data. Furthermore, the proposed framework enforces exact fulfillment of the plastic consistency condition.
To evaluate the proposed framework, a prototype model with the von Mises yield criterion and nonlinear kinematic hardening is chosen. Only 10 cycles of multiaxial proportional loading are used to generate the training data. After evaluating the proposed framework in material point simulations, we incorporate it into finite element simulations to evaluate its accuracy and computational efficiency in a boundary value problem. The results from both material point and finite element simulations show a very promising numerical performance of the neural network-based time integrator. It provides very good accuracy and numerical stability, as well as a noticeable gain in computational time for a single strain increment per load segment.
\end{abstract}


\section{Introduction}
\label{sec:introduction}

To accurately predict the mechanical response of structural components in finite element (FE) simulations, advanced nonlinear material models are essential across a wide range of engineering applications. Some examples include the investigation of rolling contact fatigue in railway rails~\cite{Lun2015, Su2019, Ghodrati2019, Talebi2025}, geotechnical and soil simulations~\cite{KARG2010}, and sheet metal forming processes~\cite{vladimirovSheetMetal2010}.
Despite their predictive capabilities, nonlinear constitutive models remain computationally costly within the FE framework, since a set of Differential Algebraic Equations (DAEs) needs to be solved in every integration point in the FE mesh, for every iteration in each time step of the simulation. Therefore, efficient material model algorithms are crucial for computational efficiency. Explicit forward Euler integration of the DAEs results in fast material models, but requires very small time steps to be numerically robust~\cite{safaei2015TI}. Implicit backward Euler integration is often preferred due to its unconditional stability, which allows for larger time steps. However, many time steps are still required per loading cycle, and a nonlinear equation system must usually be solved iteratively. The challenge is particularly severe in cyclic simulations where no stabilized behavior is obtained~\cite{leidermark2019procedures, Talebi2026}. To accelerate simulations with nonlinear material models, this study proposes a machine learning (ML) based explicit time integrator that accurately integrates the DAEs while taking much larger time steps than existing methods.

Recent studies have demonstrated that employing ML algorithms, such as Neural Networks (NNs), can reduce the computational cost of constitutive modeling, especially when integrated into FE frameworks; see, e.g., the reviews on machine learning in constitutive modeling by Fuhg et al.~\cite{fuhg2024review} and Dornheim et al.~\cite{dornheim2024NN}. In early ML-based constitutive models, NNs were trained to approximate the mapping from strain, strain history, or state variables to stress or state variable increment~\cite{hashash2004numerical, ali2019NN, HUANG2020}. In general, these approaches have low predictive accuracy beyond the training regime, as they do not account for the underlying physics~\cite{lourencco2025}. 

To address this, several approaches have been proposed in the literature to satisfy physical principles in a weak or strong form. Inspired by the idea of Physics-Informed Neural Networks (PINNs)~\cite{raissi2019PINN}, Haghighat et al.~\cite{Haghighat2023PINN} proposed a PINN-based constitutive modeling framework in which violations of elastoplastic inequality constraints are penalized in the loss function. In~\cite{FarahPIER2026}, the introduced physics-informed framework adhered to the second law of thermodynamics during training by penalizing violating analytical expressions for hypoplasticity. Weber et al.~\cite{weber2023} extended the idea of constrained NNs in~\cite{weber2021constrained} to rate-independent plasticity by enforcing physical constraints, such as material tangent symmetry and material stability, directly via error terms in the loss function. Despite their impressive ability to describe material behavior, these approaches do not guarantee physical constraints by construction. Instead, these constraints are only weakly enforced in the loss function during training. In contrast, Meyer and Ekre \cite{MeyerEkre2023} proposed a framework in which Feed Forward Neural Networks (FFNNs) are embedded directly into the evolution equations for internal variables. Independent of the training, this formulation inherently satisfies physical requirements, such as thermodynamical consistency and objectivity. Using an alternative approach, Fuhg et al.~\cite{fuhg2023modular} introduced a thermodynamically consistent framework based on isotropic and nonlinear kinematic hardening potentials. 

ML-based approaches have recently offered a promising route to accelerate the numerical integration of plasticity models, in which the local solution of nonlinear evolution equations at each integration point dominates the computational cost. The following provides elaboration on some relevant studies from the literature focused on this objective. Zhang and Mohr~\cite{zhangMohr2020} reformulated the classical return-mapping algorithm for von Mises plasticity using nonlinear isotropic hardening with explicit integration in an approximate manner. They employed an NN to estimate the tangent stiffness without imposing a priori assumptions
about the yield surface, flow rule, or hardening law. 
In~\cite{Jang2021}, during plastic loading, the stress integration algorithm was replaced with an NN in principal stress space, employing plane stress von Mises plasticity with nonlinear isotropic hardening under monotonic and cyclic loading. Further, they extended the approach to anisotropic plasticity for sheet metal forming applications~\cite {fazily2023}. Dettmer et al.~\cite{dettmer2024} presented a framework for the stress update procedure that separates the state variable update from stress prediction, using state and response NN, respectively. The NNs were trained and evaluated based on cyclic data from uniaxial elastoplasticity with linear isotropic hardening, a uniaxial elastoplastic damage model, and plane strain elastoplasticity with tensor components used as input and output features. Several PINN-type surrogate models have been investigated in literature for bypassing the need for solving the nonlinear system of equations at the material point level. Examples include the work of Eghbalian et al.~\cite{Eghbalian2023}, who proposed a PINN-based surrogate model to replace an elastoplastic material model using incremental data for monotonic loading in principal stress space. Moreover, Rezaei et al.~\cite{Rezaei2024} adopted an unsupervised PINN-based approach to train a surrogate model for accelerating the incrementation of elastoplastic material models. They evaluated the methodology on a one-dimensional plasticity model with nonlinear isotropic hardening and on a three-dimensional interface damage model for cracking behavior. In contrast to purely surrogate-based formulations, Li et al.~\cite{li2026pinnDriven} introduced a PINN-driven framework that circumvents only the local stress integration in an elastoplastic soil constitutive model under cyclic loading. Aiming to fulfill the consistency condition during plastic loading, Zhang~\cite{zhang2026Consistency} presented an unsupervised learning framework in which an NN is leveraged to predict the stress increment accurately during plastic loading in each strain increment. This objective was achieved by minimizing a loss function that incorporates the residuals of the yield function and the stress increment.
Recently, Malleval et al.~\cite{Malleval2025} proposed using NNs to approximate the solution to a scalar nonlinear constitutive equation at each integration point for an elastoviscoplastic material model. 

The core idea of this study is to propose an NN-based explicit time integration algorithm that enables robust integration over very large strain increments within a physics-encoded framework. 
The framework is designed such that loading and unloading conditions are fulfilled by construction. Training data are generated from highly resolved reference solutions obtained using a plasticity model with nonlinear kinematic hardening under multiaxial proportional loading. To ensure frame-independent predictions and reduce the dimensions of the NNs, the models are formulated with invariants as input features, combined with evolution directions inferred from the training data. Furthermore, the consistency condition during plastic loading is enforced in the framework by introducing a correction factor.
Finally, we embed the proposed framework into FE examples to assess its numerical performance and computational efficiency for boundary value problems under different strain increment sizes. 

The paper is organized as follows: In Section~\ref{sec:ML_accelerated_time_integrator}, we present the general formulation for explicit ML-based time integration of a generic rate-independent plasticity model. This is followed by the description of the adopted prototype model and the corresponding formulation of the ML-based time integration scheme in Section~\ref{sec:prototype_plasticity_model}. We explain the considered FFNNs and the training data generation strategy in Section~\ref{sec:NN_training}. Finally, in Sections~\ref{sec:results} and~\ref{sec:summary}, we discuss and summarize our findings.

\section{ML-accelerated time integration of a generic plasticity model}
\label{sec:ML_accelerated_time_integrator}

In this section, we first present the generic formulation of a thermodynamically consistent material model in a small-strain setting, followed by a description of the time integration schemes for this model. Finally, we propose an explicit ML-accelerated time integrator for solving the evolution equations of state variables.

The notation for different tensors is as follows: Second-order tensors are boldface, e.g. $\ts{t}$, fourth-order tensors are boldface and upright, e.g. $\tf{T}$. Sets of variables are written with blackboard bold, e.g. $\set{T}$. Finally, $\bullet\dev := \bullet - \tr(\bullet)\ts{I}/3$ denotes the deviatoric part of the tensor $\bullet$. 

\subsection{Generic small-strain plasticity model}
\label{subsec:generic_plasticity_formulation}

The starting point for a generic small-strain, rate-independent, plasticity model is the Helmholtz's free-energy function, $\helmholtz$, that depends on the total strain, $\eps$, and the set of $N_{\rm{s}}$ state variables, $\state = \defsetstate{\statevar}$,
\begin{align}
    \helmholtz = \helmholtz\left(\eps, \state\right)
\end{align}
The stress, $\sig$, and the (hardening) variables, $\set{A} = \defsethardening{\tn{a}}$, are defined as
\begin{align}
    \sig(\eps, \state) = \pdiff[\helmholtz]{{\eps}}, \quad 
    \tn{a}_i(\eps, \state) = -\pdiff[\helmholtz]{\statevar_i}
    \nonumber
\end{align}
Next, a yield function, $\varPhi\left(\sig, \set{A}\right)$, that defines the elastic domain $(\varPhi < 0)$ and the plastic domain $(\varPhi = 0)$, is formulated. By introducing the plastic multiplier, $\dot{\lambda}$, the Karush-Kuhn-Tucker (KKT) loading/unloading conditions are expressed as
\begin{equation}
    \varPhi \le 0, \quad \dot{\lambda} \ge 0, \quad \varPhi \, \dot{\lambda}=0
\label{eq:KKT}
\end{equation}
The state variables only evolve during plastic loading, which are described by the generic evolution laws
\begin{align}
    \dot{\statevar}_i=\dot{\lambda} \, \tn{f}_i\left(\sig(\eps, \state),\ \set{A}(\eps, \state) \right)
    \label{generic:evolution}
\end{align}

\subsection{Integration of a generic small-strain plasticity model}
\label{subsec:time_integration_generic}

In an incremental strain-controlled algorithm, the role of a material model is to calculate the current stress, $\sig$, and the updated state variables, $\state$. Here, we assume that the strain increment (from time $\oldtime{t}$ to $t$), $\Delta \eps = \eps - \oldtime{\eps}$, is linear on a material point level, i.e.,
\begin{align}
    \eps(\theta) = \oldtime{\eps} + \theta \Delta \eps
\end{align}
for $\theta \in [0, 1]$ within the time step.\footnote{For higher order time integration on the global finite element level, $\eps$ may vary non-linearly during the increment, but such higher order space-time finite element formulations are not considered herein.}  For a given $\Delta \eps$, the standard procedure is to assume an elastic response and calculate the trial stress
\begin{align}
    \sig\supscr{tr} := \sig(\eps, \oldtime{\state})
\end{align}
The assumption is true if $\varPhi\trial:= \varPhi(\sig\supscr{tr}, \oldtime{\set{A}}) < 0$, resulting in $\sig = \sig\supscr{tr}$ and $\state = \oldtime{\state}$; otherwise, the material model must solve the set of DAEs in Equations \ref{eq:KKT} and \ref{generic:evolution}. Different time integration schemes, $\tn{g}_i\supscr{TI}$, can be used to approximate the solution to Equation~\ref{generic:evolution}
\begin{align}
    \Delta \statevar_i = \tn{g}_i\supscr{TI}(\oldtime{\eps},\ \Delta\eps,\, \oldtime\state)
\end{align}
For implicit algorithms, $\tn{g}_i\supscr{TI}$ can typically not be described analytically, but requires solving a nonlinear equation system iteratively~\cite{runesson1988integration}.
In explicit algorithms, upon transition from elastic to plastic behavior, the intersection of the trial stress path with the yield surface is identified to determine the plastic part of the applied strain increment~\cite{schreyer1979accurate, owen1980FEPlasticity}. This introduces the elastic limit stress, $\sig^{\rm lim}$, as
\begin{equation}
 \sig^{\rm lim} = \sig\left(\oldtime{\eps} + s \, \Delta\eps, \oldtime{\state}\right)   
\end{equation}
where the scalar variable $s \in [0,1]$ is determined from $\varPhi(\sig^{\rm lim}, {\oldtime{\set{A}}})= 0$. 
During this elastic substep, no evolution of state variables occurs as $\varPhi < 0$. Accordingly, $\tn{g}_i\supscr{TI}$ only acts on the plastic loading part as
\begin{align}
    \Delta \statevar_i = \tn{g}_i\supscr{TI}(\oldtime{\eps}+s\Delta\eps,\ [1-s]\Delta\eps,\ \oldtime{\state})
\end{align}
For the remaining strain increment, the KKT condition (Equation~\ref{eq:KKT}) is replaced with $\varPhi = 0$, which, coupled with Equation~\ref{generic:evolution}, form the DAEs to be solved to calculate $\Delta{\statevar}_i$. The solution for the increment of the plastic multiplier, $\Delta \lambda$, is typically obtained using an iterative procedure for implicit integration, whereas it can be obtained analytically for explicit integration from $\varPhi(\sig, \set{A}) = 0$. This makes explicit time integrators numerically efficient, but they require small time steps to ensure numerical stability.

\subsection{ML-accelerated integration of a generic small-strain plasticity model}
\label{subsec:ML_based_integration_generic}

The idea of this paper is to use ML to find an explicit time integrator, $\tn{g}_{\statevar_i}\supscr{ML}$, that can handle large strain increments with good accuracy and numerical stability. Specifically, we propose to use NNs to formulate $\tn{g}_{\statevar_i}\supscr{ML}$ for each state variable, $\statevar_i$, as 
\begin{align}
    \tn{g}_{\statevar_i}\supscr{NN}(&\oldtime{\eps} + s \Delta\eps,\ [1-s]\Delta\eps,\ \oldtime{\state}) = \nonumber \\
    &\sum_{k=1}^{N} \left[
    \neuralnet_{{\statevar_i}, \,1, \,k}(\set{I}) \pdiff[I_k]{\sig\supscr{tr}} + \neuralnet_{{\statevar_i},\,2,\,k}(\set{I}) \pdiff[I_k]{\sig\supscr{lim}} + \sum_{j=1}^{N\subscr{a}} \neuralnet_{{\statevar_i},\,j+2,\,k}(\set{I})\pdiff[I_k]{\oldtime{\tn{a}}_j}\right]
    \label{eq:generic_ML_integration}
\end{align}
where $\neuralnet$ are scalar outputs of an NN, and $\set{I}= \defsetinvar{I}$ is a set of suitable invariants that depends on $\sig\supscr{tr}$, $\sig\supscr{lim}$, and $\oldtime{\tn{a}}_j$. The choice to express the NNs in invariants eliminates coordinate system dependence and lowers their dimensions.

When applying the proposed time integration scheme, the increments of the state variables are computed as 
\begin{equation}
    \Delta \statevar_i = c \, \tn{g}_{\statevar_i}\supscr{NN}
\end{equation}
where we introduce the correction factor, $c$, to ensure that $\varPhi(\sig, \set{A}) = 0$ is satisfied exactly during plastic loading.

\section{Prototype model: Plasticity model with kinematic hardening}
\label{sec:prototype_plasticity_model}

This section first presents the prototype plasticity model, which will be used to train and evaluate the proposed NN-based time integration scheme. This is followed by a detailed description of the NN-based integration.

\subsection{Prototype plasticity model}
\label{subsec:Reference_model}

Considering the generic formulation of a small-strain plasticity model explained in Section~\ref{subsec:generic_plasticity_formulation}, the prototype plasticity model assumes linear isotropic elasticity and adopts kinematic hardening. 
The set of state variables for this model is $\state = \{\ts{\epsilon}^{\rm{p}}, \, \ts{b}\}$, where $\ts{\epsilon}^{\mathrm p}$ is the plastic strain and $\ts b$ is the state variable associated with kinematic hardening. The free-energy density function, $\helmholtz$, is formulated as
\begin{align}
    \helmholtz\left(\eps, \, \state\right) =
    \frac{1}{2} \, \left[\eps - \eps\pl\right]:\tf{E}^{\mathrm{e}}:\left[\eps - \eps\pl\right]+\frac{1}{3} \, H_{\rm{kin}} \, \ts{b} : \ts{b}
\end{align}
where $H_{\rm{kin}}$ is the kinematic hardening modulus. Isotropic elasticity is considered such that the elasticity tensor, $\tf{E}^{\mathrm{e}}$, is formulated as
\begin{equation}
    \quad  \tf{E}^{\rm{e}} = 2 \, G \, \tf{I}^{\dev} + K_{\rm{b}} \, \ts{I}\otimes\ts{I} \quad \text{where} \quad K_{\mathrm{b}} = \frac{E \, G} {3\left(3 \, G-E\right)}
\end{equation}
$G$ and $E$ are the shear and Young's moduli, respectively. 
$\tf{I}^{\dev} = \tf{I} - \ts{I}\otimes\ts{I}/3$ is the 4th order deviatoric identity tensor, where $\ts{I}$ and $\tf{I}$ are the second- and fourth-order identity tensors.
The stress, $\ts{\sigma}$, and the (hardening) variables, $\set{A}=\{\sig\pl, \backstress\}$, are obtained as  
\begin{equation}
    \sig = \frac{\partial \helmholtz}{\partial \ts{\epsilon}}=\tf{E}^{\mathrm{e}}:\left[\ts{\epsilon}-\ts{\epsilon}^{\mathrm{p}}\right]
, \quad
\ts{\sigma}^{\rm p}= -\frac{\partial \helmholtz}{\partial \ts{\epsilon}^{\rm p}}=\sig, \quad
\backstress= -\frac{\partial \helmholtz}{\partial \ts{b}}=-\frac{2}{3} \, H_{\rm kin} \, \tv{b}
\end{equation}
where $\ts{\sigma}^{\rm p}$ is the plastic stress and $\backstress$ the back-stress.
The plasticity model adopts the isotropic von Mises yield function, which is formulated as
\begin{equation}
    \varPhi\left(\sig, \set{A}\right) = f\supscr{vM}\left(\sig\dev\red\right) - Y_0, \quad f\supscr{vM}\left(\sig\dev\red\right) = \sqrt{\frac{3}{2}\sig\red\dev:\sig\red\dev} \quad \text{where}\quad \sig\red\dev = {\ts{\sig}}\dev-\ts{\beta}
    \label{eq:phi_eq}
\end{equation}
$Y_0$ is the initial yield stress. The evolution of the plastic strain, $\eps\pl$, follows the associative flow rule
\begin{equation}
    \tsd{{\epsilon}}^{\rm{p}}=\dot{\lambda} \, \frac{\partial\varPhi}{\partial\ts{\sigma}} = \dot{\lambda} \, \ts{\nu} \quad \text{where} \quad \ts{\nu} = \sqrt{\frac{3}{2}} \, \frac{\sig\red\dev}{\norm{\sig\red\dev}}
    \label{eq:reference_plastic_strain}
\end{equation}
The evolution of the back-stress, $\backstress$, is assumed to follow the Armstrong-Frederick kinematic hardening law~\cite{armstrong1966mathematical} as
\begin{equation}
    \dot{\backstress}= -\frac{2}{3} \, H_{\rm{kin}} \, \dot{\ts{b}}  \quad \text{where} \quad 
    \dot{\ts{b}}=\dot{\lambda}\left[-\ts{\nu} + \frac {3} {2} \, \frac{\backstress} {{\beta}_{\infty}}\right]
    \label{eq:prototype_backstress}
\end{equation}
${{\beta}_{\infty}}$ is the saturation value of the back-stress. The material model parameter values are presented in Table~\ref{tab:material_params_chaboche}, which are inspired by those calibrated against experimental data for isotropic R260 pearlitic rail steel~\cite{Talebi2025}. The prototype material model, integrated with the implicit Backward Euler time integration algorithm to solve the set of coupled DAEs in Equations~\ref{eq:KKT},~\ref{eq:reference_plastic_strain}, and~\ref{eq:prototype_backstress}, is employed for training data generation and performance evaluation of the prototype model, integrated with the NN-based integration scheme in material point simulations; see Sections~\ref{subsec:training_data_generation} and~\ref{subsec:material_point_sim_res} respectively for further details.

\begin{table}[t!]
    \caption{Material model parameter values for the prototype plasticity model.}
    \centering \footnotesize
    \setlength{\tabcolsep}{14pt}
    \begin{tabular}{l lllll} \hline
    Material parameters& $ E $ & $G$& $Y_{0}$ & $H_{\mathrm{kin}}$ & $\beta_{\infty}$\\
    \hline
     & 200 & 75 & 400  & 150  & 500 \\
    Unit &GPa & GPa & MPa & GPa & MPa \\
    \hline
    \end{tabular}
    \label{tab:material_params_chaboche}
\end{table}

\subsection{NN-based integration of the prototype plasticity model}
\label{subsec:ML_based_integration}

\begin{figure}[!b]
    \centering
    \includegraphics[width=0.8\textwidth]{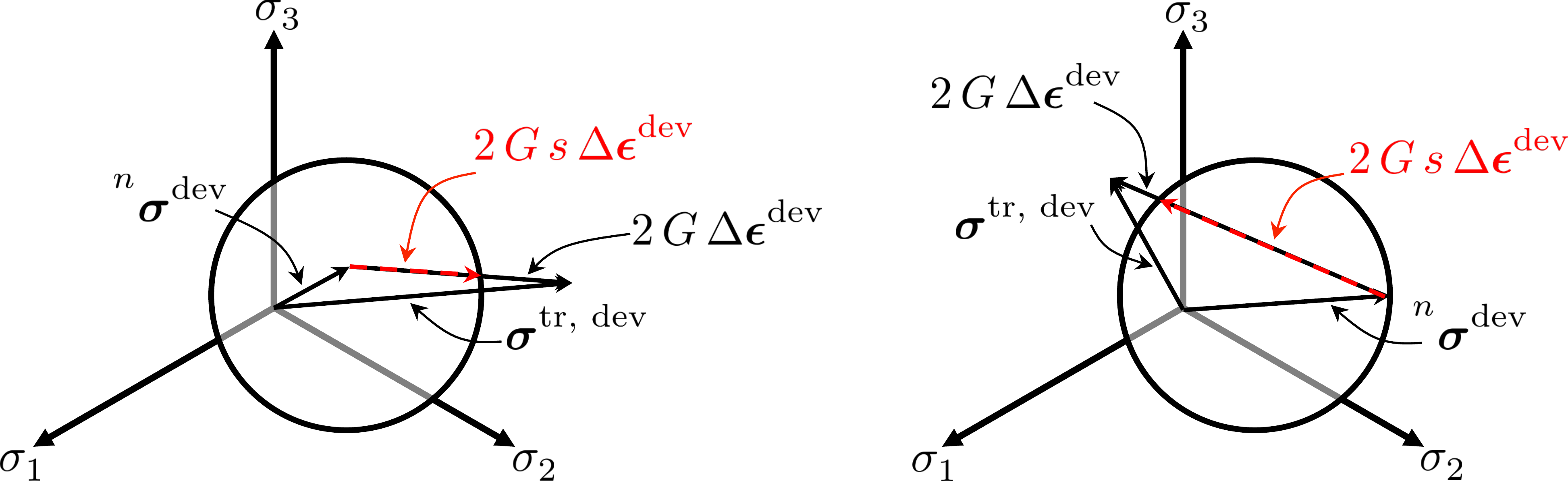}
    \caption{Schematic illustration of two cases in which the scalar variable $s$ needs to be determined. The left and right figures correspond, respectively, to the case where the old material response is elastic and to the case where trial stress path crosses the yield surface. The illustration is presented in the deviatoric stress space.}
    \label{fig:elastic_substep}
\end{figure}

When integrating the prototype plasticity model with the proposed explicit NN-based time integration scheme described in Section~\ref{subsec:ML_based_integration_generic}, for a given $\Delta\eps$, we first calculate $\varPhi\trial = \varPhi(\oldtime{\sig} + \tf{E}:\Delta\eps, \oldtime{\ts{\beta}})$. If $\varPhi\trial < 0$, the loading is elastic such that $\sig=\sig\trial$, $\eps\pl = \oldtime{\eps}\pl$, and $\ts{b} = \oldtime{\ts{b}}$. Otherwise, if $\varPhi\trial\geq0$, we need to determine if the first part of the load step is elastic. If the material response at time $\oldtime{t}$ was elastic, i.e., $\oldtime{\varPhi} = \varPhi(\oldtime{\sig}, \oldtime{\ts{\beta}}) < 0$, this will be the case as illustrated in the left part of Figure \ref{fig:elastic_substep}. However, even if $\oldtime{\varPhi} = 0$, the first part could be elastic due to a load direction change, see the right part of Figure \ref{fig:elastic_substep}. We detect this case by evaluating if $\varPhi$ decreases (i.e., becomes negative since the previous value was zero) at the beginning of the strain increment, specifically 
\begin{align}
    \left.\pdiff[{\varPhi(\oldtime{\sig} + \tf{E}:[\theta \Delta \eps], \oldtime{\ts{\beta}})}]{\theta}\right\vert_{\theta=0} \leq 0
    \label{eq:s_factor_load_dir_change}
\end{align}
If either case is satisfied, we compute the plastic part of the applied $\Delta\eps$. To be specific, the scalar variable $s\in[0,1]$ is determined such that $\varPhi(\sig^{\rm lim}, {\oldtime{\backstress}})= 0$, resulting in
\begin{align}
    &f_{s} = \frac{3}{2}\norm{\oldtime{\sig\dev} + s \, \left[\sig\trialdev - \oldtime{\sig\dev}\right] - \oldtime{\backstress}}^2 - Y_0^2 = 0
    \label{eq:s_param}
\end{align}
which can be solved analytically.

To ensure that the response of the material model, using the NN-based time integrator, is independent of the choice of coordinate system, several approaches have been proposed in literature. These include training an NN on several transformations of raw data to learn frame invariance implicitly~\cite{ling201invariance}, using a suitable set of invariants as input features~\cite{MeyerEkre2023}, or enforcing frame invariance at the neuron level~\cite{Rimoli2024symmetry}. As mentioned in Section~\ref{subsec:ML_based_integration_generic}, we adopt the second approach and choose the inputs to the NNs as invariants. This is computationally more efficient and requires less training data than the first approach~\cite{ling201invariance}. 
Following~\cite{Boehler1977}, possible choices of invariants considering the prototype model are
\begin{align}
\begin{array}{llll}
    \rm{tr}\left(\sig\trial\right), 
    & \sig\trialdev:\sig\trialdev,
    & \rm{tr}\left(\left[\sig\trialdev\right]^3\right), 
    & \\[2pt]
    \rm{tr}\left(\sig^{\rm{lim}} \right), 
    & \sig^{\rm{lim, \, dev}}:\sig^{\rm{lim, \, dev}},
    & \rm{tr}\left(\left[\sig^{\rm{lim, \, dev}} \right]^3\right), 
    &  \\[2pt]
    \rm{tr}\left(\oldtime{\backstress}\right),
    & \oldtime{\ts{\beta}}: \oldtime{\ts{\beta}},
    & \rm{tr}\left(\left[\oldtime{\ts{\beta}}\right]^3\right),
    & \\[2pt]
    \sig\trialdev : \sig^{\rm{lim, \, dev}} ,
    & \sig\trialdev : \left[\sig^{\rm{lim, \, dev}} \right]^2,
    & \left[\sig\trialdev\right]^2 : \sig^{\rm{lim, \, dev}} ,
    & \left[\sig\trialdev\right]^2 : \left[\sig^{\rm{lim, \, dev}} \right]^2, \\[2pt]
    \oldtime{\ts{\beta}} : \sig\trialdev,
    & \oldtime{\ts{\beta}} : \left[\sig\trialdev\right]^2, 
    & \left[\oldtime{\ts{\beta}}\right]^2 : \sig\trialdev,
    & \left[\oldtime{\ts{\beta}}\right]^2 : \left[\sig\trialdev\right]^2, \\[2pt]
    \sig^{\rm{lim, \, dev}} : \oldtime{\ts{\beta}},
    & \sig^{\rm{lim, \, dev}}:\left[\oldtime{\ts{\beta}}\right]^2,
    & \left[\sig^{\rm{lim, \, dev}} \right]^2 : \oldtime{\ts{\beta}},
    & \left[\sig^{\rm{lim, \, dev}} \right]^2 : \left[\oldtime{\ts{\beta}}\right]^2, \\[2pt]
    \rm{tr}\left(\sig\trialdev \cdot \sig^{\rm{lim, \, dev}}  \cdot \oldtime{\ts{\beta}}\right)
    &
    &
    &
\end{array}
\label{eq:all_invariants}
\end{align}
We restrict the nonlinearity of the inputs to the NNs to quadratic invariants; hence, the considered invariants are
\begin{align}
\begin{array}{lll}
    J_{{\rm tr}, \, {\rm tr}} =\sig\trialdev:\sig\trialdev, 
    & J_{{\rm lim}, \, {\rm lim}} = \sig^{\rm{lim,}} \, \dev:\sig^{\rm{lim,}} \, \dev, 
    & J_{\beta, \, \beta} = \oldtime{\ts{\beta}}:\oldtime{\ts{\beta}},\\[2pt]
    J_{{\rm tr}, \, {\rm lim}}   = \sig\trialdev:\sig^{\rm{lim,}} \, \dev, 
    & J_{\beta, \, {\rm tr}} = \oldtime{\ts{\beta}}:\sig\trialdev, 
    & J_{{\rm lim}, \, \beta} = \sig^{\rm{lim,}} \, \dev:\oldtime{\ts{\beta}}
\end{array}
\end{align}
and these are collected in the set $\set{J}$. 

For training of the NNs, it is vital that inputs and outputs are scaled appropriately. With the chosen set of invariants, we obtain the following formulation for the scaled NN-based time integrators, $\hat{\tn{g}}_{\tn{s}_i}^\mathrm{NN}$,
\begin{align}
\hat{\tn{g}}_{\epsilon^{\rm p}}\supscr{NN}&=
\neuralnet_{\, \eps\pl, \, 1}\left(\hat{\set{J}}\right) \, \hat{\sig}\trialdev +  \neuralnet_{\, \eps\pl, \, 2}
\left(\hat{\set{J}}\right)  \, \hat{\sig}^{\rm{lim,}} \, \dev + \neuralnet_{\, \eps\pl, \, 3}
\left(\hat{\set{J}}\right) \, {\oldtime{\hat{\backstress}} } \label{eq:g_hat_eps_p}
\\
\hat{\tn{g}}_{b}\supscr{NN}&=
\neuralnet_{\, b, \, 1}\left(\hat{\set{J}}\right) \, \hat{\sig}\trialdev +  \neuralnet_{\, b, \, 2}
\left(\hat{\set{J}}\right)  \, \hat{\sig}^{\rm{lim,}} \, \dev + \neuralnet_{\, b, \, 3}
\left(\hat{\set{J}}\right) \, {\oldtime{\hat{\backstress}} } \label{eq:g_hat_beta}
\end{align}
with the unscaled time integrators,
\begin{align}
    {\tn{g}}_{\epsilon^{\rm p}}\supscr{NN}= \mathtt{SF}_{\epsilon^{\rm{p}}}\, \hat{\tn{g}}_{\epsilon^{\rm p}}\supscr{NN}, \quad 
    {\tn{g}}_{b}\supscr{NN}=  \mathtt{SF}_{b}\, \hat{\tn{g}}_{b}\supscr{NN}
    \label{eq:unscaled_g}
\end{align}
All scaled variables are denoted by a hat, $\hat{\bullet}$, and are scaled by the values in the training data: all invariants in $\set{J}$ are individually normalized by the range found in the training data, and the evolution directions ${\sig\trialdev}$, $\sig^{\rm{lim,}} \, \dev$, $\oldtime{{\backstress}}  $ are individually normalized by their maximum Frobenius norms found in the training data. Moreover, scaling factors $\mathtt{SF}_{\epsilon^{\rm{p}}}$ and $\mathtt{SF}_{b}$ are also calculated using maximum Frobenius norms of ${\Delta\eps\pl}$ and $\Delta \ts{b}$ from the training data. 

Finally, the increments of the state variables are calculated as
\begin{equation}
     \Delta\eps\pl = c \, {\tn{g}}_{\epsilon^{\rm p}}\supscr{NN} \;\; \text{and} \;\;
     \Delta\ts{b} = c \, {\tn{g}}_{b}\supscr{NN}
     \label{eq:statevar_incr_final}
 \end{equation}
 where the correction factor, $c$, is computed from $\varPhi(\sig, \set{A}) = 0$. This leads to the expression
\begin{align}
    &f_c=\frac{3}{2}
    \norm{\sig\trialdev - \oldtime{\backstress} - c \, \left[ 2\, G \,{\tn{g}}_{\epsilon^{\rm p}}\supscr{NN} -  2/3 \, H_{\rm kin}\,{\tn{g}}_{b}\supscr{NN} \right]}^2
    - Y_0^2 = 0 
    \label{eq:c_factor}
\end{align}
where we also adopt $\Delta \backstress=-2/3 \, H_{\rm kin} \, \Delta \ts{b}$ from Equation~\ref{eq:prototype_backstress}.
We choose the solution of $c$ from Equation~\ref{eq:c_factor} that is closest to unity.
Note that, during the training of the neural networks, $c=1$ is assumed.
The overall numerical algorithm based on the proposed NN-based integration scheme is presented in Algorithm~\ref{alg:algorithm}. 

\begin{algorithm}[t!]
\setstretch{1.3} 
\caption{Numerical algorithm considering the prototype material model integrated by the proposed NN-based time integrator for a given $\Delta\eps$}
\label{alg:algorithm}
Compute $\varPhi\trial$ \\
\eIf{ $\varPhi\trial < 0$}{
    Elastic state \\
        $\sig=\sig\trial = \oldtime{\sig} + {\tf{E}}^{\mathrm{e}}:\Delta\eps$ \\
        $\ts\eps\pl = \oldtime{\eps\pl}$ \\
        $\backstress = \oldtime{\backstress}$\\
}{
Plastic state \\
    \eIf{$\oldtime{\varPhi} < 0 \,\,$ \text{\rm{or}} \,\, $\left.\pdiff[{\varPhi(\oldtime{\sig} + \tf{E}:[\theta \Delta \eps], \oldtime{\ts{\beta}})}]{\theta}\right\vert_{\theta=0} \leq 0$}{
    
        Determine $s$ from Equation~\ref{eq:s_param}\\
        Compute $\sig^{\rm{lim}}$\\
        \Indp
            $\sig^{\rm{lim}} = \oldtime{\sig} + {\tf{E}}^{\mathrm{e}}: s \, \Delta\eps$ \\
        \Indm
    }{
    $\sig^{\rm{lim}} = \oldtime{\sig}$
    }
  {  Compute ${\tn{g}}_{\epsilon^{\rm p}}\supscr{NN}$ and ${\tn{g}}_{b}\supscr{NN}$ \\
    Determine $c$ from Equation~\ref{eq:c_factor}\\
    Update the state variables  and stresses\\
    \Indp
        $\eps\pl = \oldtime{\eps\pl} + c \, {\tn{g}}_{\epsilon^{\rm p}}\supscr{NN}$ \\
        $\backstress = \oldtime{\backstress} - \frac{2}{3} \, H_{\rm{kin}} \, c \, {\tn{g}}_{b}\supscr{NN}$ \\
        $\sig=\sig^{\rm{lim}} + {\tf{E}}^{\mathrm{e}}:\left[\Delta\eps - \Delta\eps\pl\right]$ \\
    \Indm
}
}
\end{algorithm}

\subsection{Evaluation of the considered evolution directions}
\label{subsec:select_evolution_directions}

The chosen six invariants in $\set{J}$ result in three evolution directions, ${{\sig}\trialdev}$, ${\sig}^{\rm{lim,}} \, \dev$ and $\oldtime{{{\backstress}}}$; see Equation~\ref{eq:generic_ML_integration}. By analyzing the training data, the goal of this section is to determine if (a) these directions span all evolution directions and are sufficient for the model, and (b) if the number of considered directions can be reduced.

\begin{figure}[!t]
    \centering
    \includegraphics[width=0.88\textwidth]{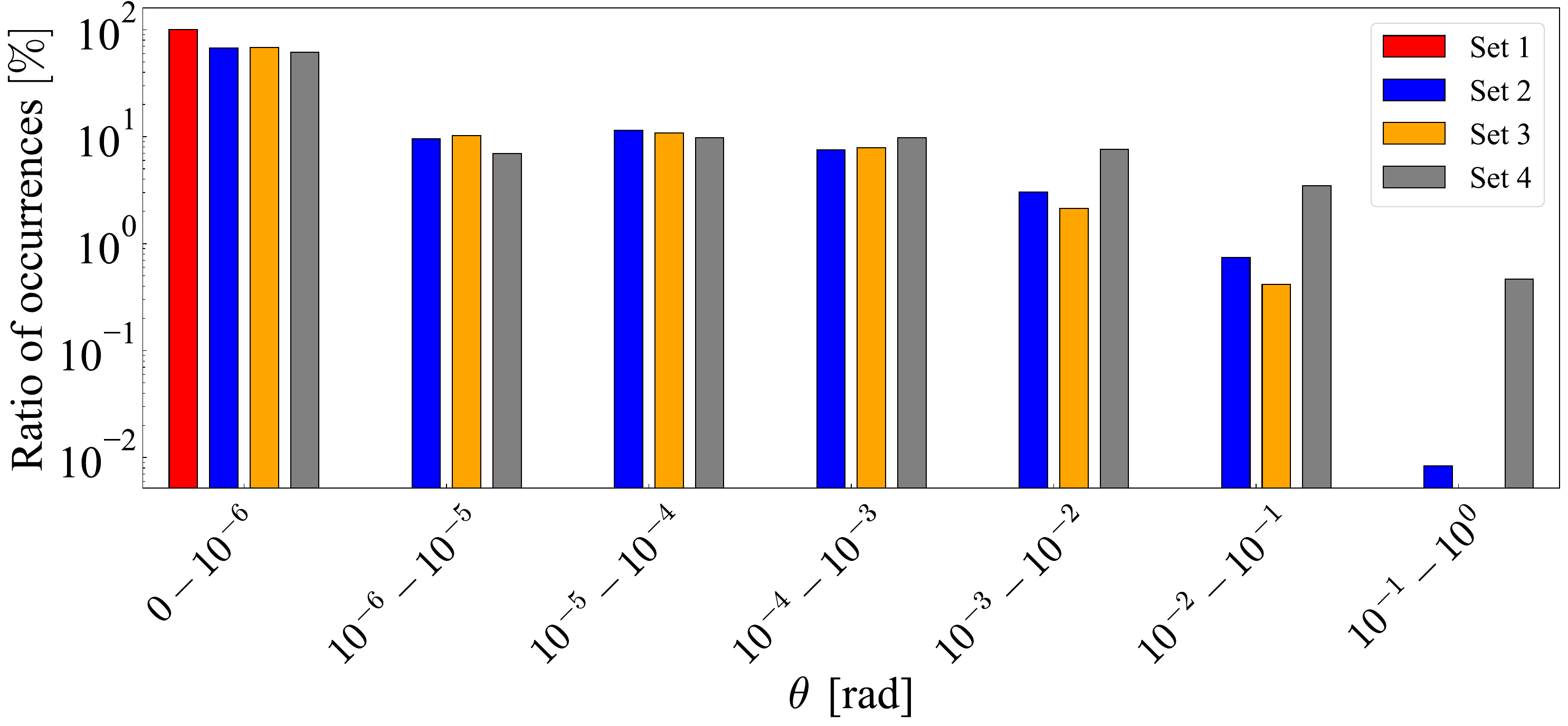}
    \caption{The distribution of the computed angles, $\theta$, between the target and predicted $\Delta\eps\pl$ considering all training data for four sets of evolution directions: set 1: $\left\{{\sig}\trialdev, {\sig}^{\rm{lim, \ dev}},\oldtime{{\backstress}}\right\}$, set 2: $\left\{{\sig}\trialdev, {\sig}^{\rm{lim, \, dev}}\right\}$, set 3: $\left\{{\sig}\trialdev, \oldtime{{\backstress}}\right\}$, and set 4: $\left\{{\sig}^{\rm{lim, \, dev}}, \oldtime{{\backstress}}\right\}$.}
    \label{fig:angles}
\end{figure}

For each training data, we have given ${\Delta\statevar}_i^{\mathrm{target}}$, i.e., $\Delta\eps\pl$ and $\Delta\ts{b}$, as well as  ${{\sig}}\trialdev$, ${{\sig}}^{\rm{lim, \ dev}}$ and $\oldtime{{\backstress}}$. Considering Equations~\ref{eq:g_hat_eps_p}-~\ref{eq:statevar_incr_final}, we can determine coefficients $\alpha_{1i}$ - $\alpha_{3i}$ by a least square fit
\begin{equation}
    \text{min} \; \norm{ {\Delta\statevar}_i^{\mathrm{target}} - \mathtt{SF}_{\statevar_i} \left[ \, \alpha_{1i} \, {\hat{\sig}}\trialdev + \alpha_{2i} \,
    {\hat{\sig}}^{\rm{lim, \ dev}} +
    \alpha_{3i} \, \oldtime{\hat{\backstress}}\right]}^2
\end{equation}
With the identified coefficients, the predicted increments are obtained as
\begin{equation}
{\Delta\statevar}_i^{\mathrm{pred}}=\mathtt{SF}_{\statevar_i} \left[ \, \alpha_{1i} \, {\hat{\sig}}\trialdev + \alpha_{2i} \,
    {\hat{\sig}}^{\rm{lim, \ dev}} +
    \alpha_{3i} \, \oldtime{\hat{\backstress}}\right]
\end{equation}
To evaluate if the training data can be described by the chosen evolution directions, we compute the angular error, $\theta_i$, as
\begin{equation}
\theta_i = \cos^{-1}\left( \frac{ {\Delta\statevar}_i^{\mathrm{target}} : {\Delta\statevar}_i^{\mathrm{pred}} }   { \norm{{\Delta\statevar_i}^{\mathrm{target}}} \norm{{\Delta\statevar_i}^{\mathrm{pred}}} }\right)
\end{equation}
If $\theta_i$ is close to 0, then the selected evolution directions can accurately express the training data; otherwise, additional directions should be introduced.
We have compared the resulting $\theta_i$ values for four sets of evolution directions. Specifically, sets $1$-$4$ are $\left\{{\sig}\trialdev, {\sig}^{\rm{lim,}} \, \dev, \oldtime{{\backstress}}\right\}$, $\left\{\ts\sig\trialdev, {\sig}^{\rm{lim,}} \, \dev\right\}$, $\left\{\ts\sig\trialdev, \oldtime{\backstress}\right\}$, and $\left\{{\sig}^{\rm{lim,}} \, \dev, \oldtime{\backstress}\right\}$. The computed $\theta_i$  values in terms of relative number of occurrences considering $\Delta\eps\pl$ are illustrated in Figure~\ref{fig:angles}, showing that using all the three directions (set 1) is both necessary and sufficient to describe the evolution of $\eps\pl$. The same conclusion applies to $\ts{b}$, for which the angular errors are shown in Appendix~\ref{appendix:additional_results}. Table~\ref{tab:theta_values_statevar} summarizes the maximum angular errors, where the maximum errors for set 1 are comparable to the numerical precision.

\section{Neural networks and training}
\label{sec:NN_training}

This section describes the NN model employed in the study to formulate $\tn{g}_{\statevar_i}\supscr{NN}$, the approach adopted to generate training data, and the training results.

\begin{table}[t!]
    \caption{Maximum values of $\theta$ between the target and predicted $\Delta\eps\pl$ and $\Delta \ts{b}$ considering four sets of evolution directions. The evaluated sets are: set 1: $\left\{{\sig}\trialdev, {\sig}^{\rm{lim, \, dev}},\oldtime{{\backstress}}\right\}$, set 2: $\left\{{\sig}\trialdev, {\sig}^{\rm{lim, \, dev}}\right\}$, set 3: $\left\{{\sig}\trialdev, \oldtime{{\backstress}}\right\}$, and set 4: $\left\{{\sig}^{\rm{lim, \, dev}}, \oldtime{{\backstress}}\right\}$.}
    \centering \footnotesize
    \setlength{\tabcolsep}{12pt}
    \vspace{0.5\baselineskip}
    \begin{tabular}{l llll} \hline
    Sets & Set 1 &  Set 2 &  Set 3 &  Set 4\\
    \hline
    
    $\theta$ (for $\Delta\eps\pl$)& 
    $2.98 \times 10^{-8}$ & $1.00 \times 10^{-1}$ & $2.56 \times 10^{-2}$  & $1.48 \times 10^{-1}$ \\
    $\theta$ (for $\Delta \ts{b}$) & 
    $2.98 \times 10^{-8}$ & $1.38 \times 10^{-1}$ & $1.60 \times 10^{-2}$ & $1.11\times 10^{-1}$ \\
    \hline
    \end{tabular}
    \label{tab:theta_values_statevar}
\end{table}

\subsection{Feed-Forward Neural Networks (FFNNs)}
\label{subsec:NN}

We use fully connected FFNNs to formulate ${\tn{g}}_{\statevar_i}\supscr{NN}$ through 
$\neuralnet_{\, \statevar_i, \, 1}$ - $\neuralnet_{\, \statevar_i, \, 3}$ for each state variable, $\statevar_i$. 
A fully connected FFNN establishes a mapping from an input vector, in our case $\hat{\set{J}}$, to an output vector, in our case $\neuralnet_{\, \statevar_i, \, 1}$ - $\neuralnet_{\, \statevar_i, \, 3}$, by propagating information through $m$ hidden layers. Each hidden layer $l$ computes its output $\underline{\ts{x}}_l$ using an activation function $a_l$ as $\underline{\ts{x}}_l = a_{l}\left(\underline{\underline{\ts{W}}}_l \, \underline{\ts{x}}_{l-1} + \underline{\ts{b}}_{l}\right)$,
where $\underline{\underline{\ts{W}}}_l$ and $\underline{\ts{b}}_{l}$ are trainable weight matrices and bias vectors associated with layer $l$. In this study, the activation function for all hidden layers is chosen as $a_{l} = {\rm{tanh}}(x)$.
The considered FFNNs consist of $m = 8$ hidden layers with 8 neurons and are trained over 30000 epochs. The FFNNs' weights and biases are randomly initialized before being optimized with the gradient-based Rectified Adam (RAdam) optimizer~\cite{liu2019RAdam} in the PyTorch library~\cite{paszke2019pytorch}, using a learning rate of $10^{-4}$. Although the selected hyperparameters were determined through trial and error to provide satisfactory training performance, further improvements in training efficiency may be possible through automated hyperparameter optimization; see~\cite{feurer2019hyperparameter}.

To facilitate the optimization of the network parameters, i.e., weights and biases, during training~\cite{gorji2020RNN}, and to prevent vanishing gradients caused by large input values to the activation function, ${\rm{tanh}}(x)$~\cite{rosenkranz2023normalization}, scaling has been introduced as described in Section~\ref{subsec:ML_based_integration}. 
\begin{figure}[!t]
    \centering
    \includegraphics[width=0.7\textwidth]{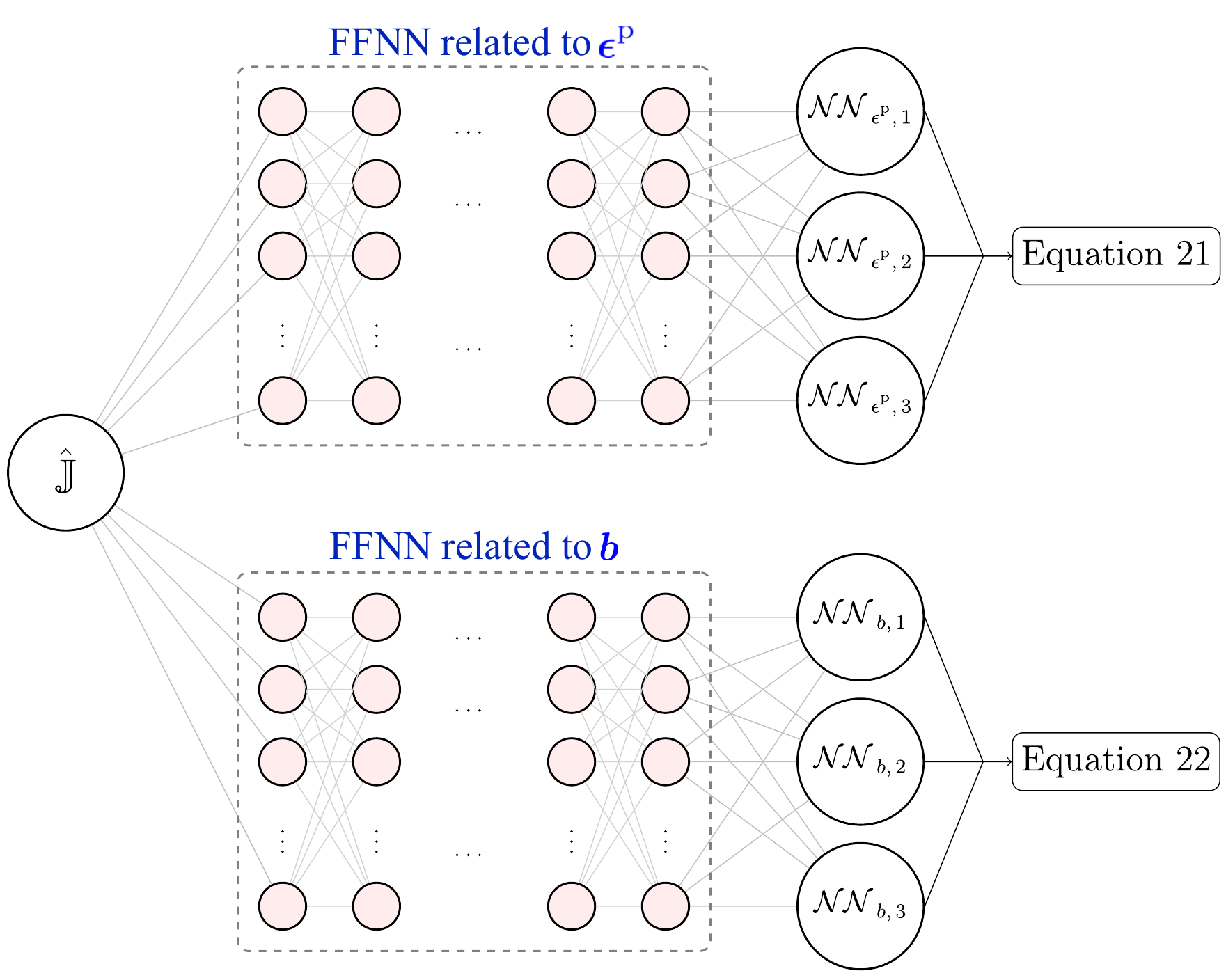}
    \caption{Schematic illustration of the inputs and outputs of the NNs, which are embedded in Equations~\ref{eq:g_hat_eps_p} and~\ref{eq:g_hat_beta}.}
    \label{fig:NN_architecture}
\end{figure}
In Equations~\ref{eq:g_hat_eps_p} and ~\ref{eq:g_hat_beta}, we use one FFNN to formulate the increment of each state variable, $\Delta\statevar_i$, as illustrated in Figure~\ref{fig:NN_architecture}. Each FFNN produces three scalar outputs denoted by $\neuralnet_{\statevar_i,1}-\neuralnet_{\statevar_i,3}$. The reason for considering separate FFNNs rather than a single FFNN with six outputs is to keep the networks smaller, thereby improving training efficiency and better training performance.
After using the outputs of the FFNNs in Equations~\ref{eq:g_hat_eps_p} and~\ref{eq:g_hat_beta} to predict $\hat{\tn{g}}_{\statevar_i}\supscr{NN}$, the mean squared error type loss, ${\mathcal{L}_{\statevar_i}}$, can be calculated as
\begin{align}
    {\mathcal{L}_{\statevar_i}}
    &= \frac{1}{N_{\mathrm{data}} } \sum_{j=1}^{N_{\mathrm{data}}} \left[\left[{{\Delta \hat{\statevar}_i}} \right]_{j}^{\rm{target}} - 
    \hat{\tn{g}}_{\statevar_i}\supscr{NN}
    \right]^2 +  \lambda_{\mathrm{p}} \, \norm{\ts p_{\statevar_i}}_2^2 
    \label{eq:loss_function}
\end{align}
Here, $N_{\mathrm{data}}$ denotes the total number of training data samples, and $\left[{{\Delta \hat{\statevar}_i}} \right]^{\rm{target}}$ is the target state variable increments, normalized with $\mathtt{SF}_{\statevar_i}$.
We apply L2 regularization to enhance the generalizability of the FFNNs~\cite{herrmann2024deepLearning, Brunton_Kutz_2019_NN}. This is done by augmenting the loss function with a differentiable penalty term corresponding to the square of the L2-norm of the network parameters $\ts p$, where $\lambda_{\rm{p}} = 10^{-6}$. 

\subsection{Training data generation}
\label{subsec:training_data_generation}

\begin{figure*}[!t]
    \centering
        \begin{subfigure}{1.\textwidth}
        \centering
        \includegraphics[width=0.92\textwidth]{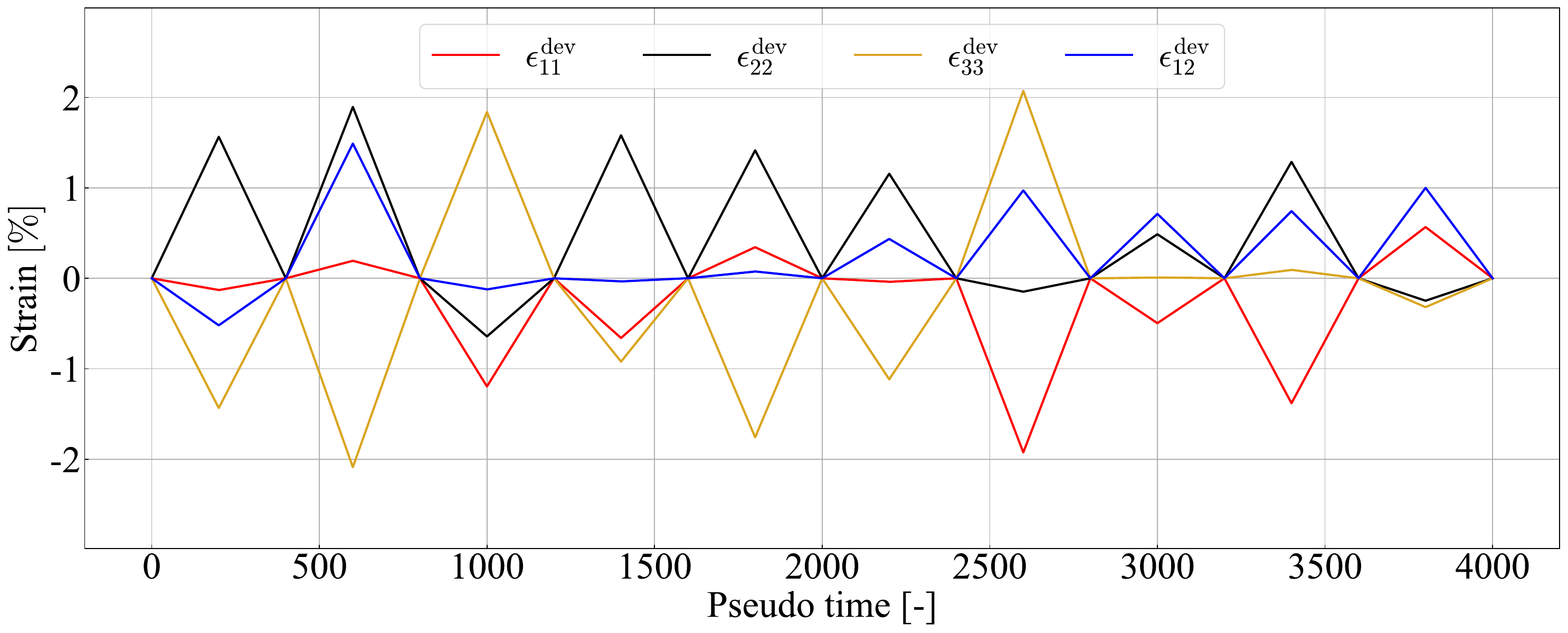}
    	\caption{}
    	\label{fig:strain_path}
        \vspace{0.8\baselineskip}
    \end{subfigure}   
    \begin{subfigure}{1.\textwidth}
    \centering
        \includegraphics[width=0.5\textwidth]{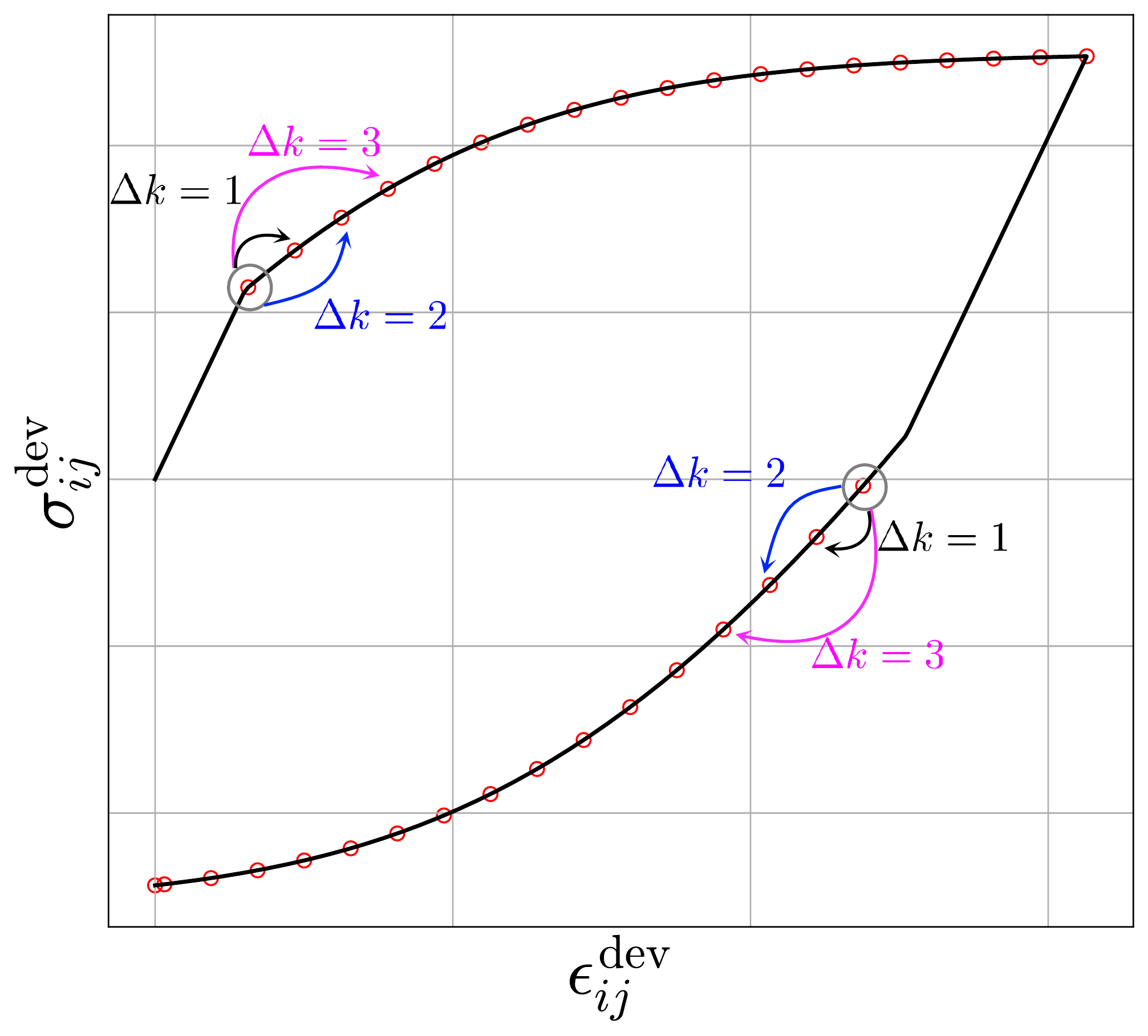}
        \caption{}
        \label{fig:data_generation}
    \end{subfigure}
\caption{(a) Deviatoric multiaxial pulsating strain path for 10 loading cycles and (b) schematic illustration of data generation approach after excluding the data points with elastic material behavior.}
\label{fig:training_data_details}
\end{figure*}

To train the FFNNs, we generate data using the prototype plasticity model presented in Section~\ref{subsec:Reference_model}, integrated with the implicit backward Euler time integration scheme. As mentioned in Section~\ref{subsec:time_integration_generic}, on a material point level, we assume a linear variation of the strain increment within the considered time step. 
A single deviatoric multiaxial pulsating strain path is generated over 10 loading cycles, each discretized into 400 time steps; see Figure~\ref{fig:strain_path}. 
A random peak value, $\norm{\eps\dev}\in[0.0,\ 3.5]\%$, is generated for each cycle. 
The choice of pulsating loading is motivated by the application of railway mechanics, where rails are subjected to purely pulsating loading~\cite{Xinli2016, bernasconi2006multiaxial, Talebi2026}. The strain histories are used in material point simulations, and the resulting stress-strain cycles are employed to generate training data samples. 

We adopt an approach similar to that described in~\cite{weber2023} to generate training data, which is schematically presented in Figure~\ref{fig:data_generation}. We first split the data into load segments with linear strain paths, which, in this case, correspond to half-cycles. Next, data points corresponding to elastic material behavior, i.e., those for which $\varPhi\trial < 0$, are filtered out. As mentioned in Section~\ref{subsec:ML_based_integration}, during the simulations, it is first assessed whether the scalar variable $s$ needs to be computed. Thus, the data points marked with a gray circle in Figure~\ref{fig:data_generation} are the first points used for training data generation in each load segment. For a load segment with $N_{\rm{p}}$ points, $k-1$ data samples are generated for each point, $k \in [2, N_{\rm{p}}]$, which results in a total of $ N_{\rm{p}} \, (N_{\rm p}-1)/2$ data samples.  One data sample for the increment size $\Delta k \in [1, \, k-1]$ consists of
\begin{equation}
\begin{aligned}
    &\sig\trialdev, \, \sig^{\rm{lim, \, dev}}, \oldtime{\backstress}  \,  \quad \text{where} \quad n = \, k-\Delta k \\
    &\set{J} = \lbrace J_{{\rm tr}, \, {\rm tr}}, \, J_{{\rm lim}, \, {\rm lim}}, \, J_{\beta, \, \beta}, \, J_{{\rm tr}, \, {\rm lim}}, \, J_{\beta, \, {\rm tr}}, \, J_{{\rm lim}, \, \beta} \rbrace \\
    &\Delta\eps\pl, \, \Delta\ts{b}\
\end{aligned}
\end{equation}
Since the aim of the NN-based integration scheme is to accelerate cyclic simulations by allowing large $\Delta\eps$, we have not used all of the possible data points in each cycle for data generation. Instead, the data points are taken every 5 increments, which also results in more efficient training. The training data generation strategy has produced 12012 training data samples.

\subsection{FFNN training}
\label{subsec:model_training}

\begin{figure*}[!t]
    \centering
    \begin{subfigure}{0.49\textwidth}
        \includegraphics[width=1\textwidth]{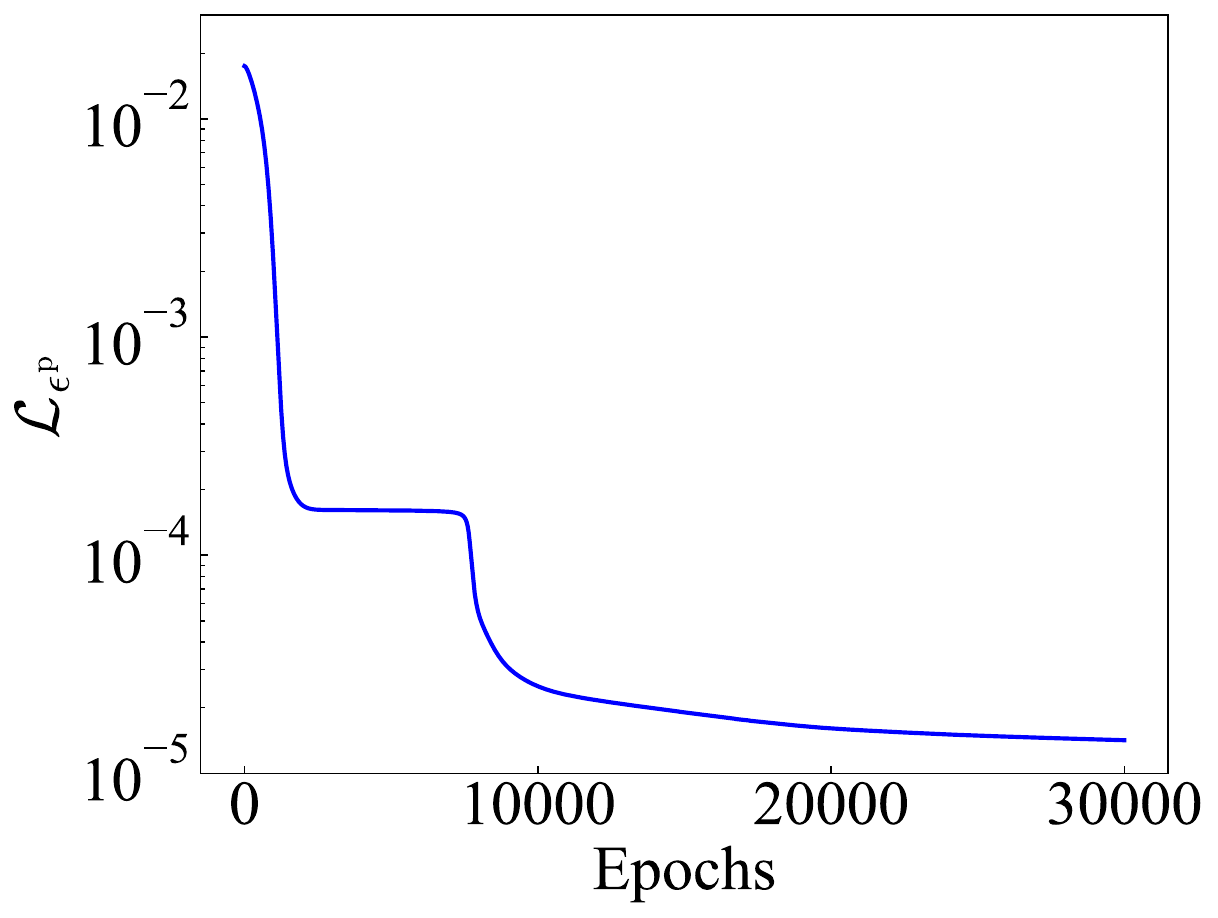}
        \caption{}
        \label{fig:loss_delta_eps_p}
    \end{subfigure}
    \begin{subfigure}{0.49\textwidth}
        \includegraphics[width=1\textwidth]{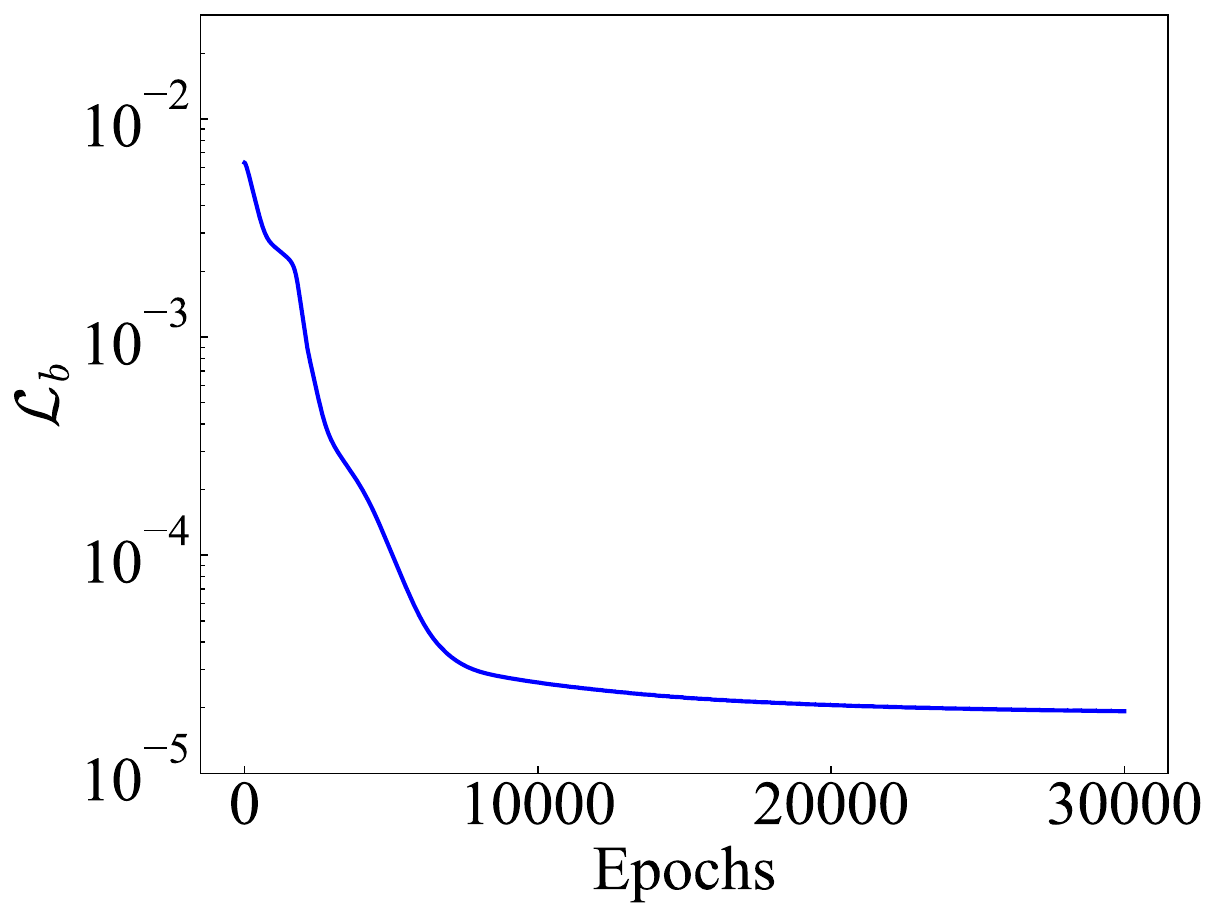}
	\caption{}
	\label{fig:loss_delta_beta}
    \end{subfigure} 
\caption{Loss function evolutions considering (a) ${\mathcal{L}_{\epsilon^{\mathrm{p}}}}$ and (b) ${\mathcal{L}_b}$ during training.}
\label{fig:loss_evolutions}
\end{figure*}

The evolutions of the loss functions, ${\mathcal{L}_{\epsilon^{\mathrm{p}}}}$ and ${\mathcal{L}_{b}}$, during the training procedure are shown in Figure~\ref{fig:loss_evolutions}. In the last epoch, ${\mathcal{L}_{\epsilon^{\mathrm{p}}}}$ and ${\mathcal{L}_{b}}$ reach their lowest values of $1.42 \times 10^{-5}$ and $1.93\times 10^{-5}$, respectively. More training did not result in a significant reduction in the training losses.

\begin{figure*}[!t]
    \centering
    \begin{subfigure}{0.55\textwidth}
        \includegraphics[width=1\textwidth]{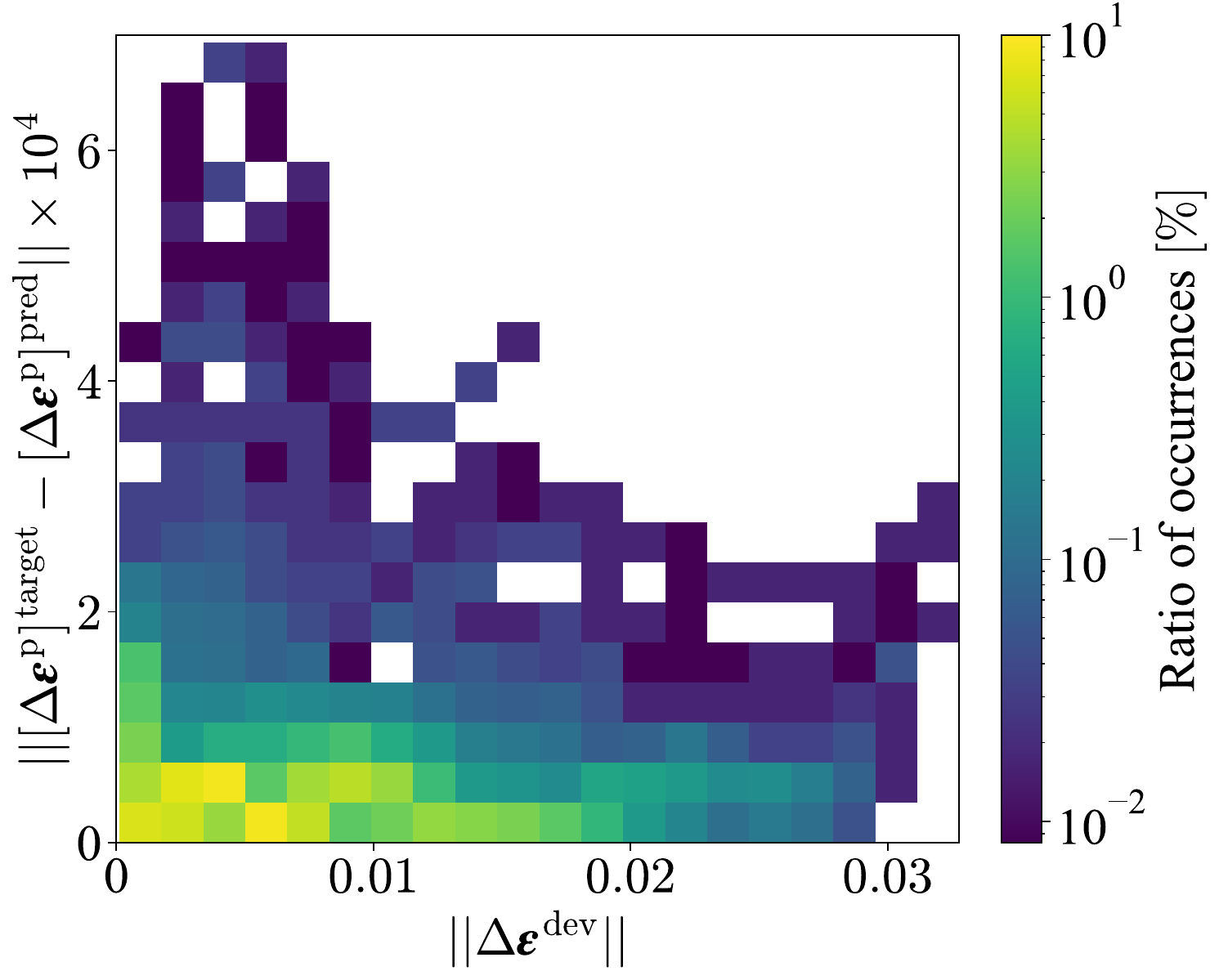}
        \caption{}
        \label{fig:delta_eps_p_pred}
    \end{subfigure}
    \hspace*{0.005\textwidth} 
    \begin{subfigure}{0.55\textwidth}
        \includegraphics[width=1\textwidth]{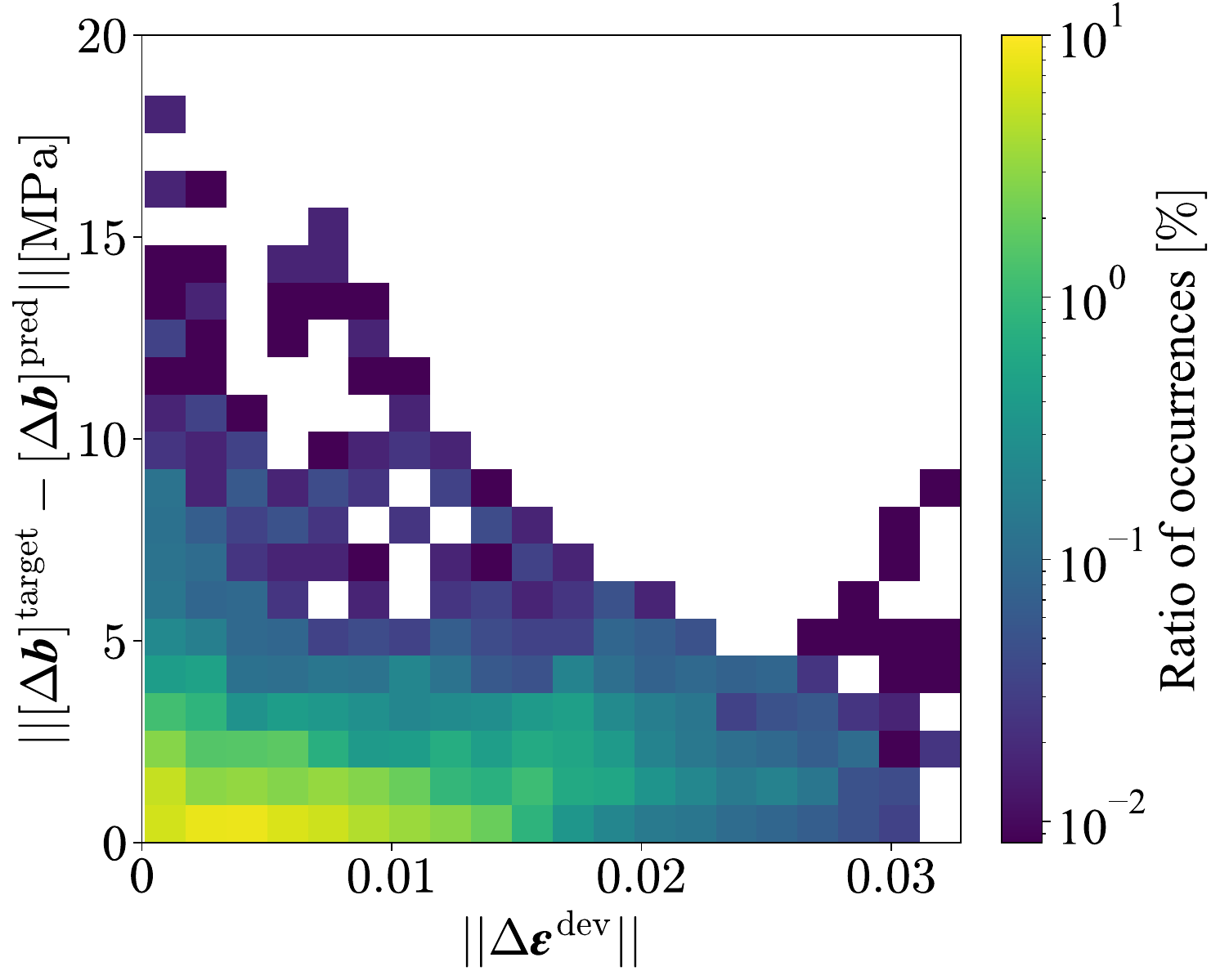}
	\caption{}
	\label{fig:delta_beta_pred}
    \end{subfigure} 
\caption{Two-dimensional histograms of the norm of absolute errors in (a) $\Delta\eps\pl$ and (b) $\Delta \ts{b}$ versus $\norm{\Delta\eps\dev}$, considering the training data. White bins correspond to zero occurrences.}
\label{fig:error_predictions}
\end{figure*}

Figures~\ref{fig:delta_eps_p_pred} and~\ref{fig:delta_beta_pred} present the distribution of the norms of absolute errors in $\Delta\eps\pl$ and $\Delta\ts{b}$ for all training data with different strain increment sizes. The training data generation strategy produces more data samples for small strain increments than for large ones. However, the largest errors for both $\Delta\eps\pl$ and $\Delta\ts{b}$ are observed for $\norm{\Delta\eps\dev} < 0.01$, although they occur with low frequency. As $\norm{\Delta\eps\dev}$ increases, the maximum error decreases. This trend breaks for $\norm{\Delta\eps\dev} > 0.03$, where there is less training data, and the accuracy of the trained FFNNs decreases.

\section{Results and discussions}
\label{sec:results}

In this section, we present and discuss the results from the performance evaluation of the proposed NN-based time integration. In particular, we have employed the NN-based integration scheme in material point and FE simulations and assessed its performance in terms of accuracy and computational time. In the following, for both simulations, the reference model denotes the model that uses the prototype material model, integrated with the implicit backward Euler time integration algorithm. By using very small time steps, the reference model provides the true solution to the DAEs that are being solved. The NN-based model refers to the model that also adopts the prototype material model, but is integrated with the proposed NN-based integration. The NN-based model has been implemented according to Algorithm~\ref{alg:algorithm}, with the trained FFNNs embedded into Equations~\ref{eq:g_hat_eps_p} and~\ref{eq:g_hat_beta}.

\subsection{Material point simulations}
\label{subsec:material_point_sim_res}

\begin{figure*}[!t]
    \centering
    \begin{subfigure}{0.47\textwidth}
        \includegraphics[width=1\textwidth]{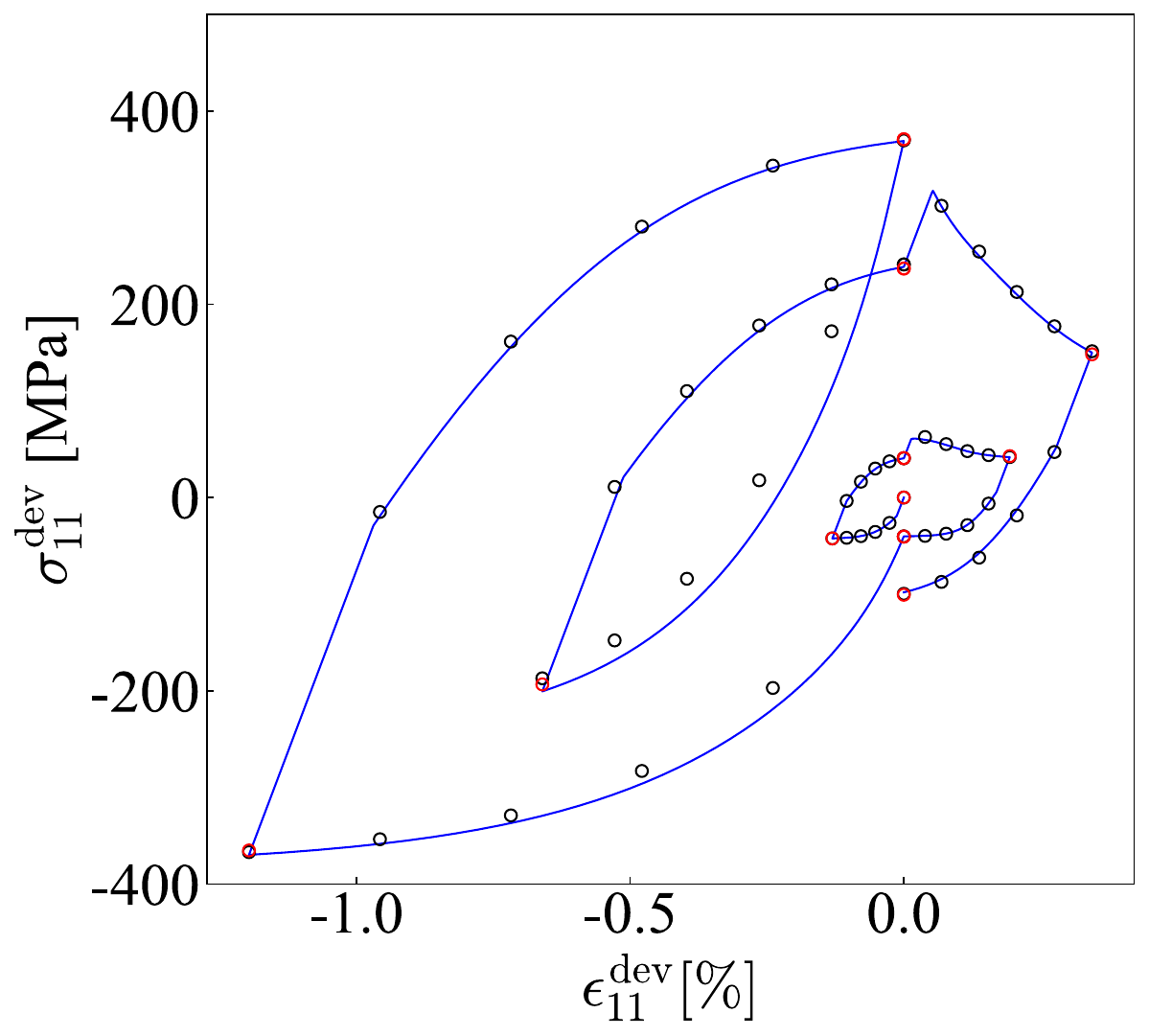}
        \vspace{0.05\baselineskip}
	\end{subfigure}
    \hspace*{0.03\textwidth} 
    \begin{subfigure}{0.47\textwidth}
        \includegraphics[width=1\textwidth]{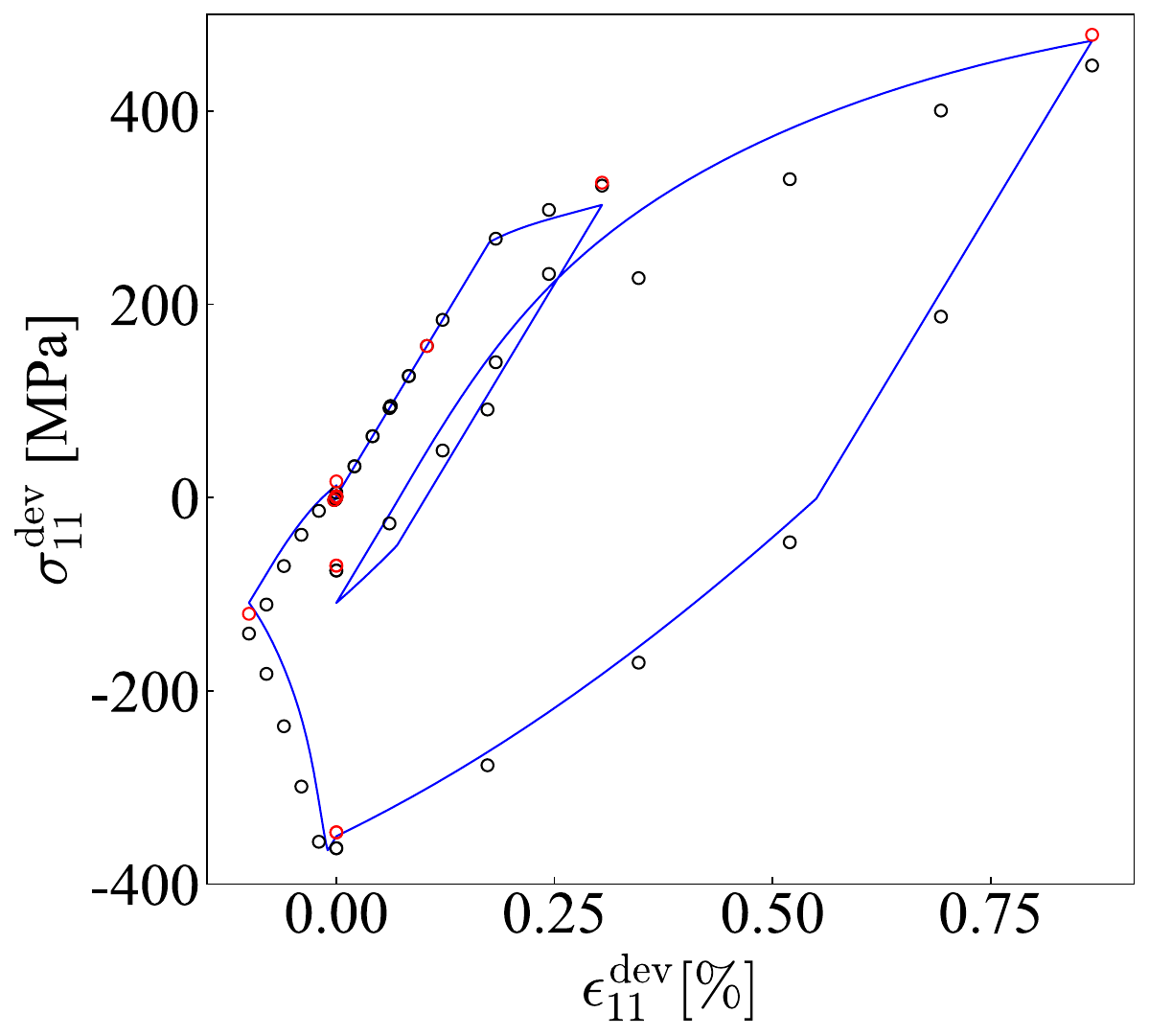}
        \vspace{0.05\baselineskip}
	\end{subfigure}
    \begin{subfigure}{0.47\textwidth}
        \includegraphics[width=1\textwidth]{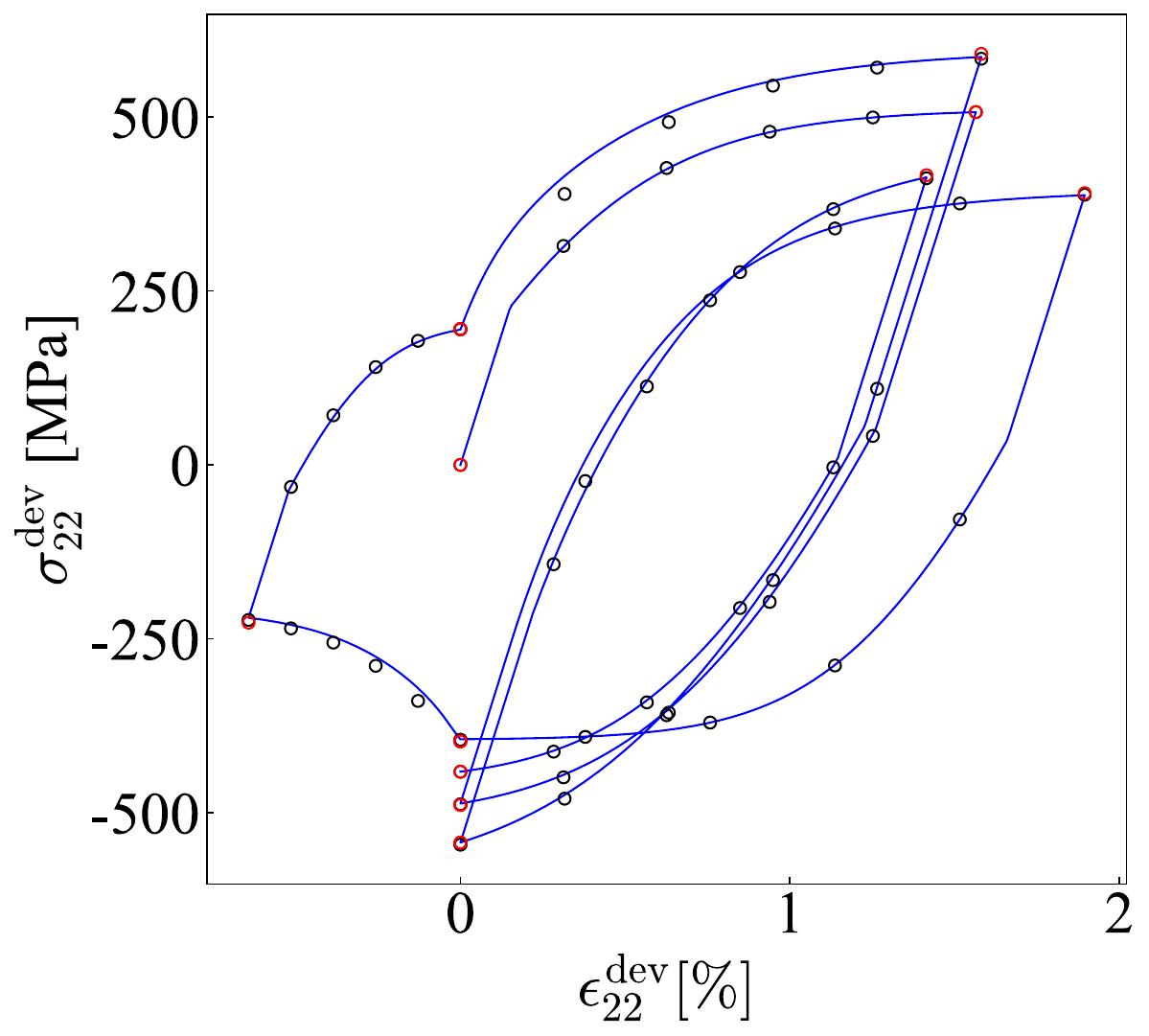}
        \caption{Training data}
        \label{fig:stress_strain_com_22_training_all}
	\end{subfigure}
    \hspace*{0.03\textwidth} 
    \begin{subfigure}{0.47\textwidth}
        \includegraphics[width=1\textwidth]{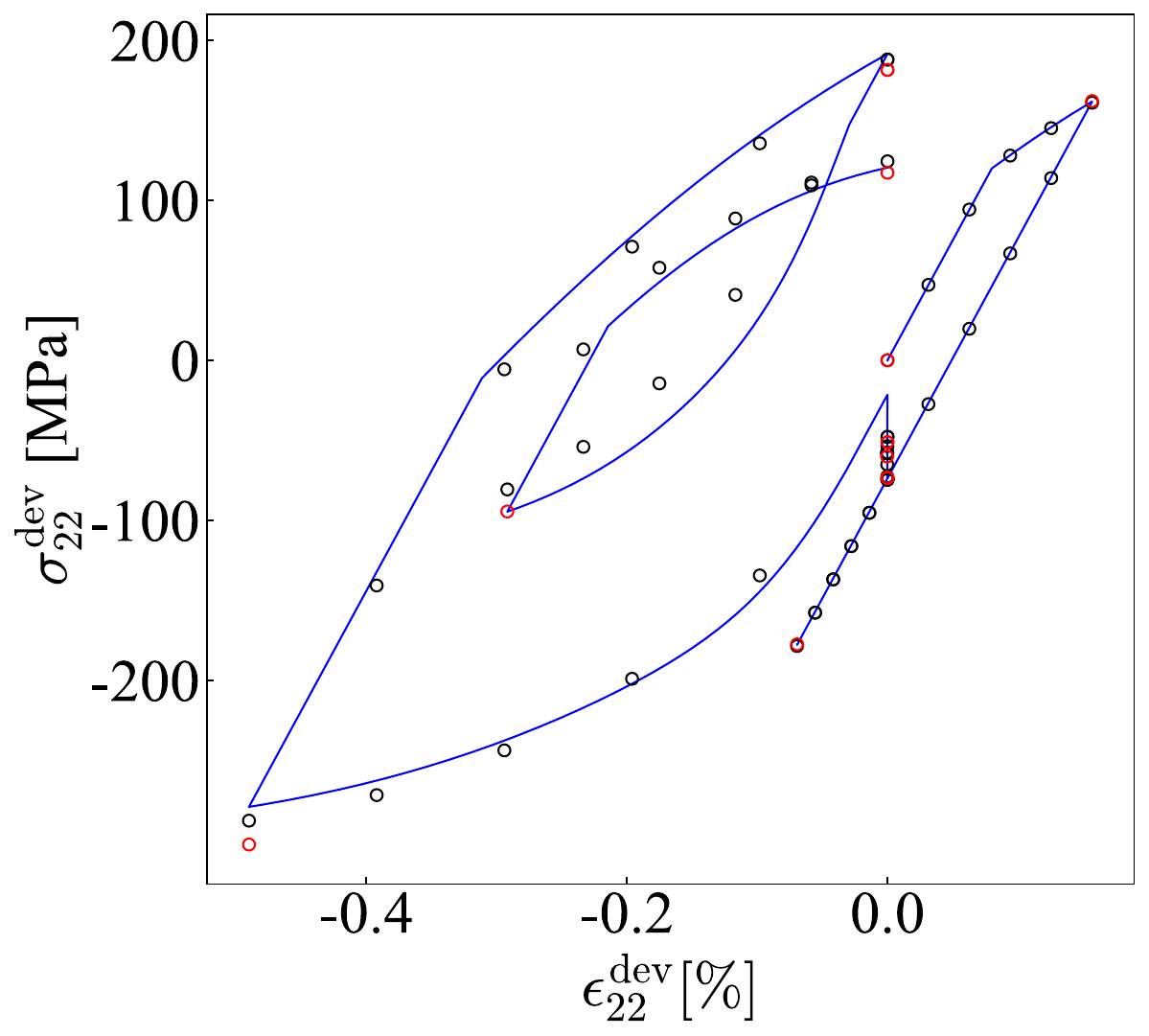}
        \caption{Test data}
         \label{fig:stress_strain_com_22_test_all}
	\end{subfigure}
\caption{The axial stress-strain responses from the reference material model (blue) and the NN-based model. The black and red circles correspond to the results from the NN-based model with 5 and 1 strain increments per half-cycle, respectively. Only the first 5 cycles are shown.}
\label{fig:stress_strain_cycles_materil_sim}
\end{figure*}

Figure~\ref{fig:stress_strain_com_22_training_all} presents the deviatoric axial responses over the first 5 cycles for the reference and NN-based models, considering the material point simulation training data. The predictions of the NN-based model, using 5 strain increments and even a single increment per half-cycle, are in close agreement with those of the reference model, in which each half-cycle is discretized into 200 increments. 

\begin{figure}[!t]
    \centering
    \includegraphics[width=1.01\textwidth]{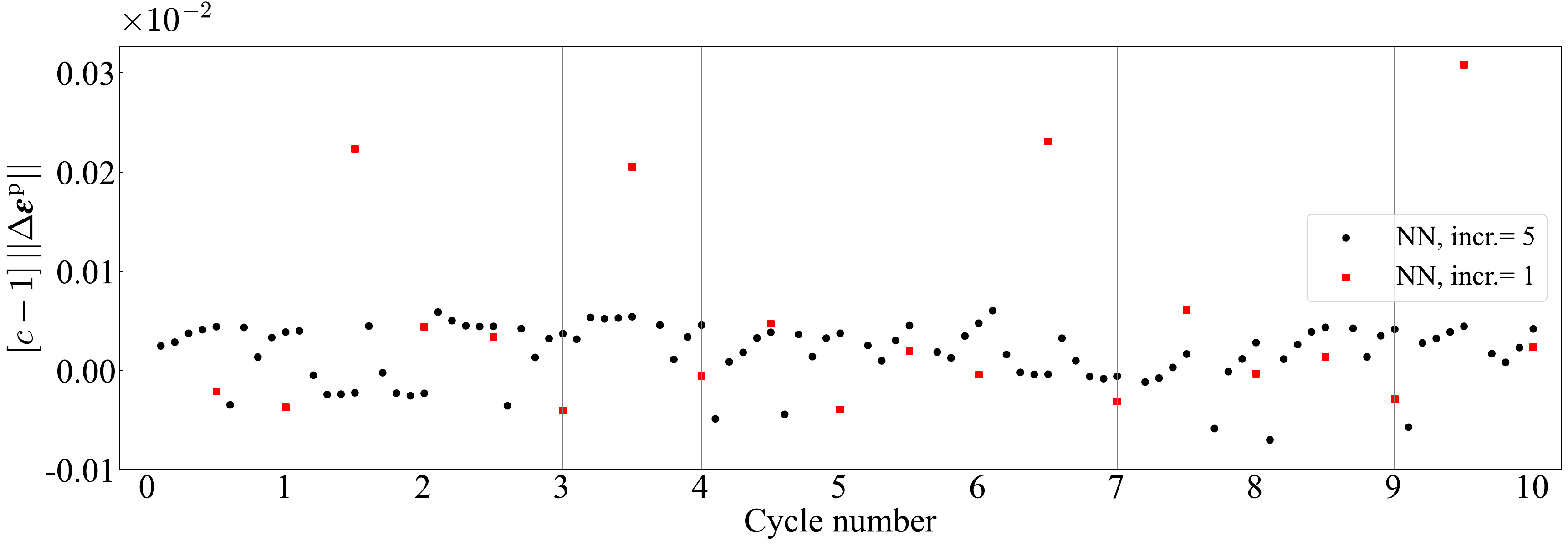}
    \caption{Calculated errors in $\norm{\Delta\eps\pl}$ due to assuming $c=1$, i.e., not enforcing the consistency condition, in each cycle when taking 5 and 1 strain increments per half-cycle (incr.), using the training data. NN refers to the NN-based model. The vertical dashed lines indicate the boundaries between consecutive cycles.}
    \label{fig:c_factor_training}
\end{figure}

To investigate the influence of not accounting for the correction factor, $c$, i.e., $c=1$ during the simulations, on the predictions of the NN-based model, we have calculated the errors in $\norm{\Delta\eps\pl}$ as $[c-1] \, \norm{\Delta\eps\pl}$ and presented them in Figure~\ref{fig:c_factor_training}. Considering the case with 5 strain increments per half-cycle, the $c$ factors have not been calculated in every strain increment within a cycle, since some increments correspond to elastic loading/unloading, e.g., in cycle 4. In contrast, in the case with 1 increment per half-cycle, they have been computed in every increment. The errors arising from the assumption of $c=1$ during plastic loading are more significant in the case using 1 increment with larger $\norm{\Delta\eps\pl}$.
\begin{figure*}[!t]
    \centering
    \includegraphics[width=0.99\textwidth]{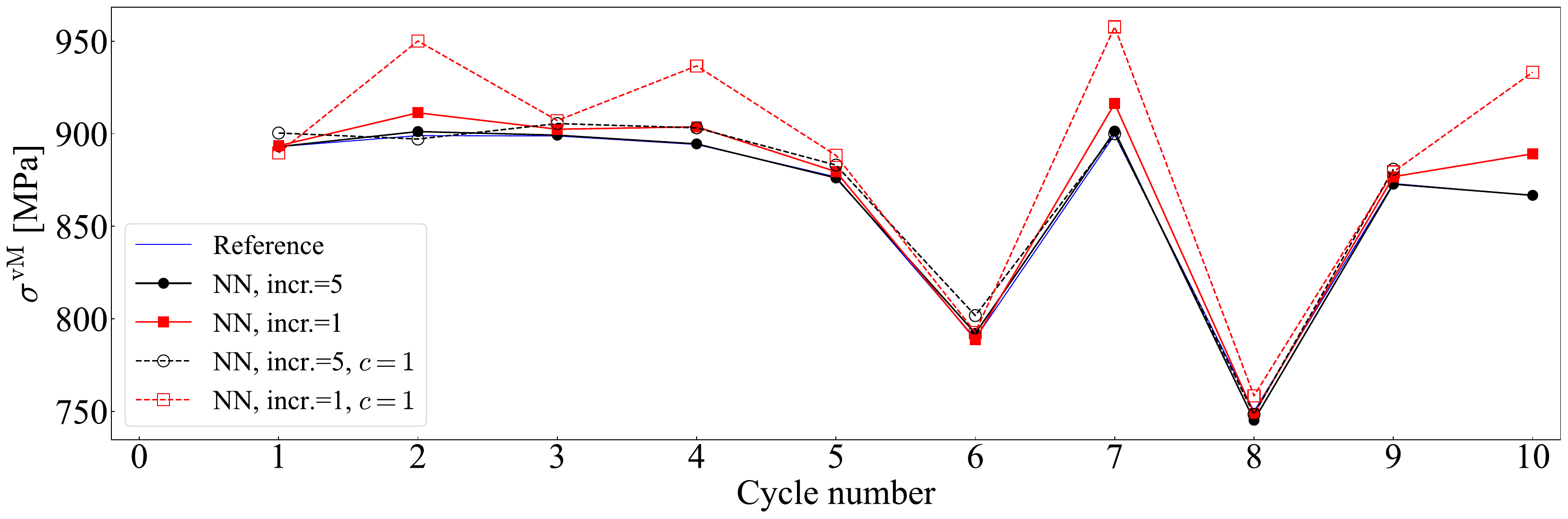}
    \caption{von Mises stresses at the peak strains over 10 cycles considering the training data. NN refers to the NN-based model, and incr. denotes the number of increments per half-cycle. $c=1$ means that the consistency condition has not been enforced.}
    \label{fig:vonMises_stress_training}
\end{figure*}
The consequences of this assumption in Equation~\ref{eq:statevar_incr_final} are further shown in Figure~\ref{fig:vonMises_stress_training}, which compares the von Mises stress values, $\sigma^{\rm{vM}}$, at the peak strain of each cycle from the reference and NN-based models. It should be noted that not enforcing the consistency condition, $\varPhi=0$, during plastic loading might lead to the calculation of the $s$ variable in the next step if $\oldtime{\varPhi} < 0$; see Section~\ref{subsec:ML_based_integration}. In the last cycle for the case with 5 increments per half-cycle, $\oldtime{\varPhi} >0$, which is inconsistent. Although Equation~\ref{eq:s_factor_load_dir_change} was fulfilled, no solution for the $s$ variable using Equation~\ref{eq:s_param} was found.
These results highlight the importance of the $c$ factor for our proposed NN-based time integration framework. Hereafter, the discussion is based on the results obtained when $c$ is calculated during plastic loading. 

Considering Figure~\ref{fig:vonMises_stress_training}, the simulation of 10 cycles has resulted in a maximum absolute error of 4.8 MPa for the NN-based model with 5 strain increments per half-cycle and 22.6 MPa with 1 increment. Given that the maximum $\sigma^{\rm{vM}}$ predicted by the reference model is approximately 900 MPa, the results demonstrate a very good predictive capability of the NN-based model. 
Moreover, despite the NN-based time integration being explicit, we note that it provides good accuracy for very large strain increments. Additional results for the shear responses are provided in Appendix~\ref{appendix:additional_results}. 
\begin{figure}[!t]
    \centering
    \includegraphics[width=1\textwidth]{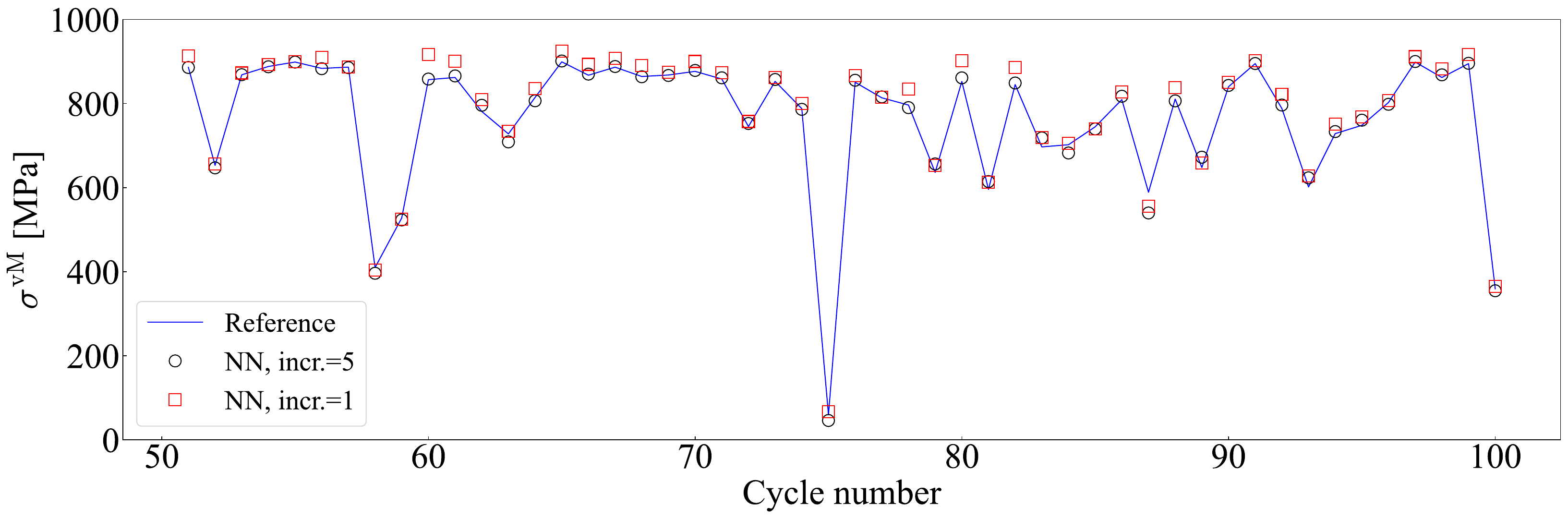}
    \caption{von Mises stresses at the peak strains for the test data (the last 50 of 100 simulated cycles are shown).}
    \label{fig:vM_stress_vs_cycle_test_material_sim}
\end{figure}

\begin{figure*}[!t]
    \centering
    \includegraphics[width=0.98\textwidth]{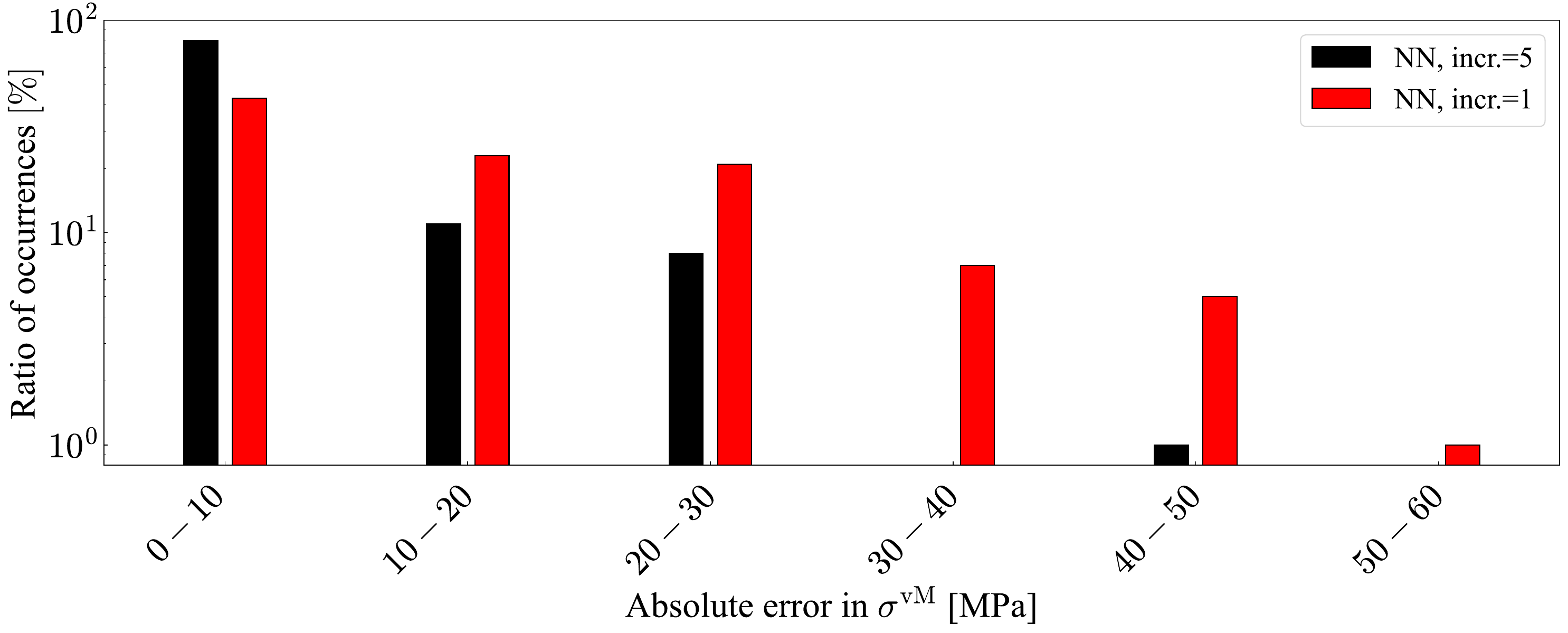}
    \caption{absolute error distribution in $\sig^{\rm{vM}}$, when comparing the results at the peak strains from the NN-based model with those from the reference model. incr. denotes the number of increments per half-cycle.}
    \label{fig:hist_vonMises_stress_error_material_sim}
\end{figure*}

To evaluate the extrapolation performance of the NN-based model, we have simulated 100 cycles with randomly generated variable strain ranges. Figure~\ref{fig:stress_strain_com_22_test_all} shows an overall good fit with the reference results considering both 5 and 1 strain increments per half-cycle. This good agreement can also be observed in Figure~\ref{fig:vM_stress_vs_cycle_test_material_sim}, presenting the $\sigma^{\rm{vM}}$ values at the peak strains in each cycle. 
To be specific, the maximum absolute errors in $\sigma^{\rm{vM}}$ over 100 cycles are roughly 49.3 MPa and 60.0 MPa for 5 and 1 strain increments per half-cycle, respectively, relative to the maximum $\sigma^{\rm{vM}}$ of about 900 MPa.
The distribution of absolute errors in $\sigma^{\rm{vM}}$ at the peak strains over 100 cycles, presented in Figure~\ref{fig:hist_vonMises_stress_error_material_sim}, shows that most errors are distributed within the lowest interval of 0-10 MPa. Despite that $6\%$ of the cases considering 1 increment have resulted in absolute errors larger than 40 MPa compared to $1\%$ for 5 increment cases, both show promising predictions.

\subsection{FE simulations}
\label{subsec:FE_simulations}

\begin{figure}[!t]
    \centering
    \includegraphics[width=1\textwidth]{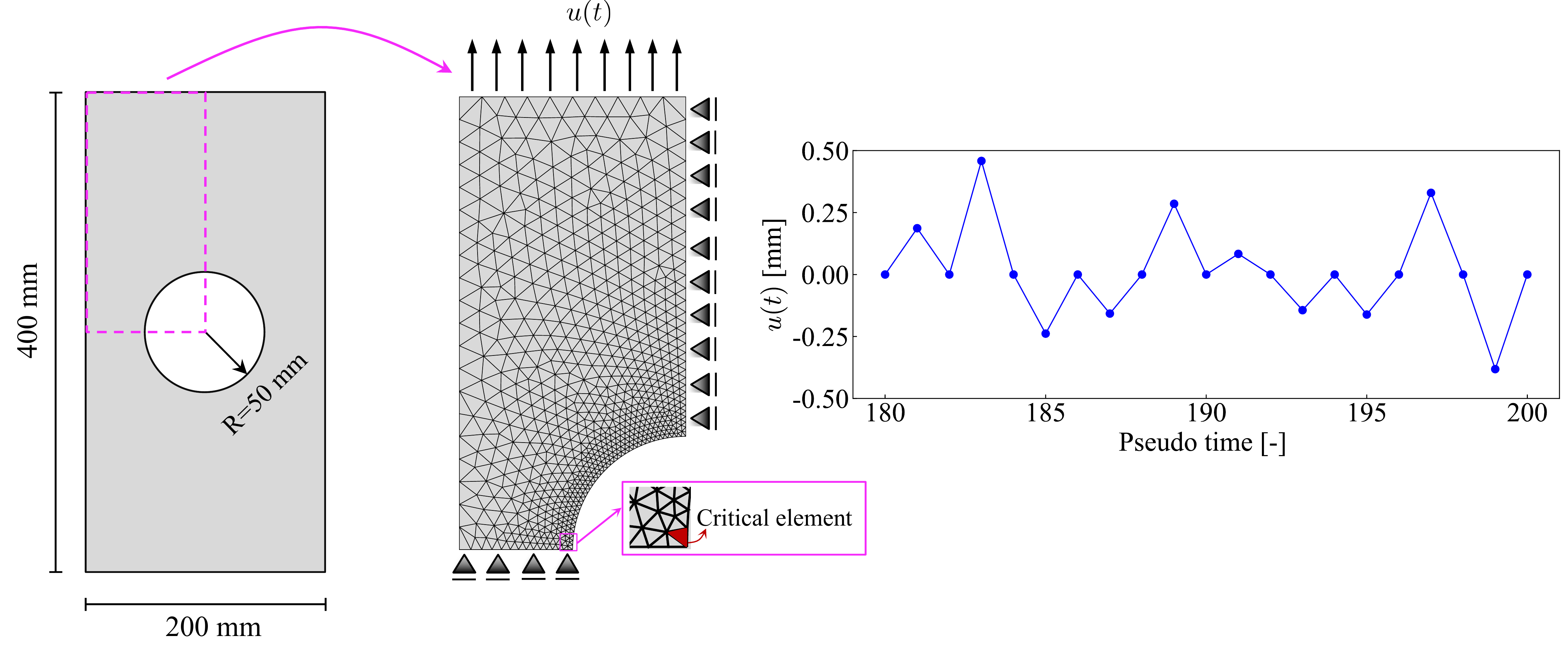}
    \caption{Geometry and boundary condition for the plate with a hole under displacement-controlled pulsating loading. The last 10 of 100 simulated cycles are shown.}
    \label{fig:FE_setup}
\end{figure}

To demonstrate the applicability of the NN-based time integration in boundary value problems, we have incorporated it into the commercial FE code Abaqus~\cite{abaqus2025}. The FE model is a quarter of a rectangular plate with a central circular hole. The plate geometry and considered pulsating displacement-control loading with varying ranges are shown in Figure~\ref{fig:FE_setup}. The FE mesh consists of 1362 linear plane strain triangular elements. The plate is subjected to 100 loading cycles, during which the vertical displacement is prescribed along the upper edge of the quarter model. Considering the reference FE model, each half-cycle is discretized into 100 displacement increments over the pseudo time interval from 0 to 1. The results from the NN-based FE model, using 5 and 1 displacement increments per half-cycle, are compared against those from the reference FE model. The material models have been implemented in Abaqus as user-defined subroutines, and for the NN-based model, the displacement iterations have been performed using the algorithmic tangent stiffness presented in Appendix~\ref{appendix:ats_tensor}. For both the reference and NN-based FE models, line search was enabled to aid convergence during the displacement iterations.

\begin{figure}
    \centering
    \includegraphics[width=0.95\textwidth]{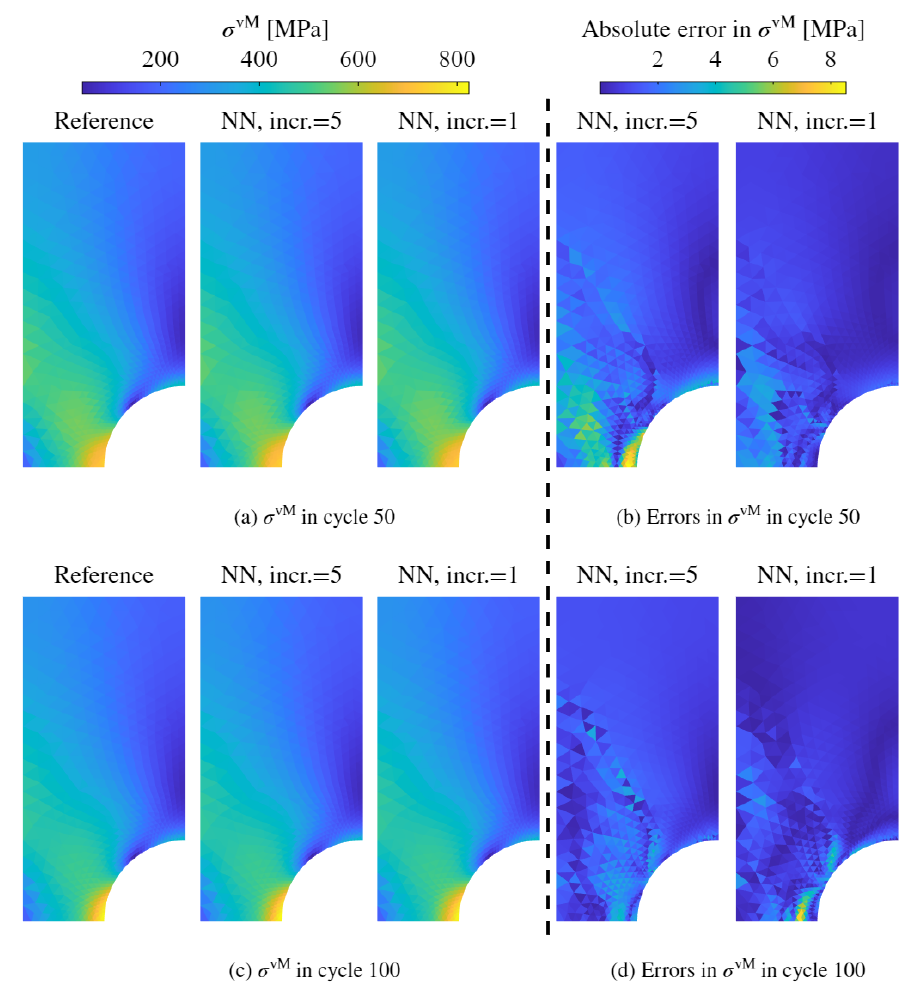}
    \caption{von Mises stress, $\sigma^{\rm{vM}}$, and the corresponding absolute error distributions in cycle 50 and 100, presented at the peak displacement. NN refers to the NN-based FE model with the number of displacement increments per half-cycle indicated by incr.}
    \label{fig:vonMises_stress_ref_ML_combined}
\end{figure}

\begin{figure*}[!t]
    \centering
        \includegraphics[width=1\textwidth]{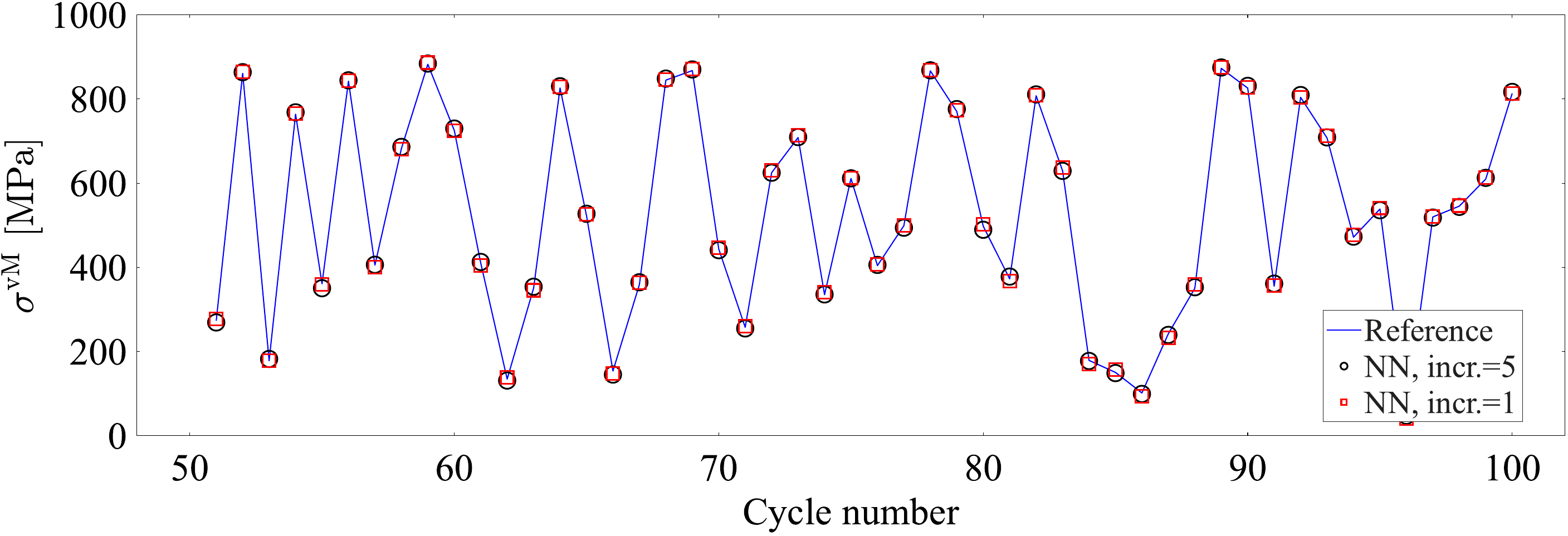}
    \caption{von Mises stresses, $\sigma^{\rm{vM}}$, in the critical element at the peak displacements, predicted by the reference and the NN-based FE models. Results are shown for the last 50 of 100 simulated cycles. incr. denotes the number of increments per half-cycle.}
\label{fig:FE_vonMises_el79}
\end{figure*}

\begin{figure*}[!h]
    \centering
    \begin{subfigure}{0.48\textwidth}
        \includegraphics[width=1\textwidth]{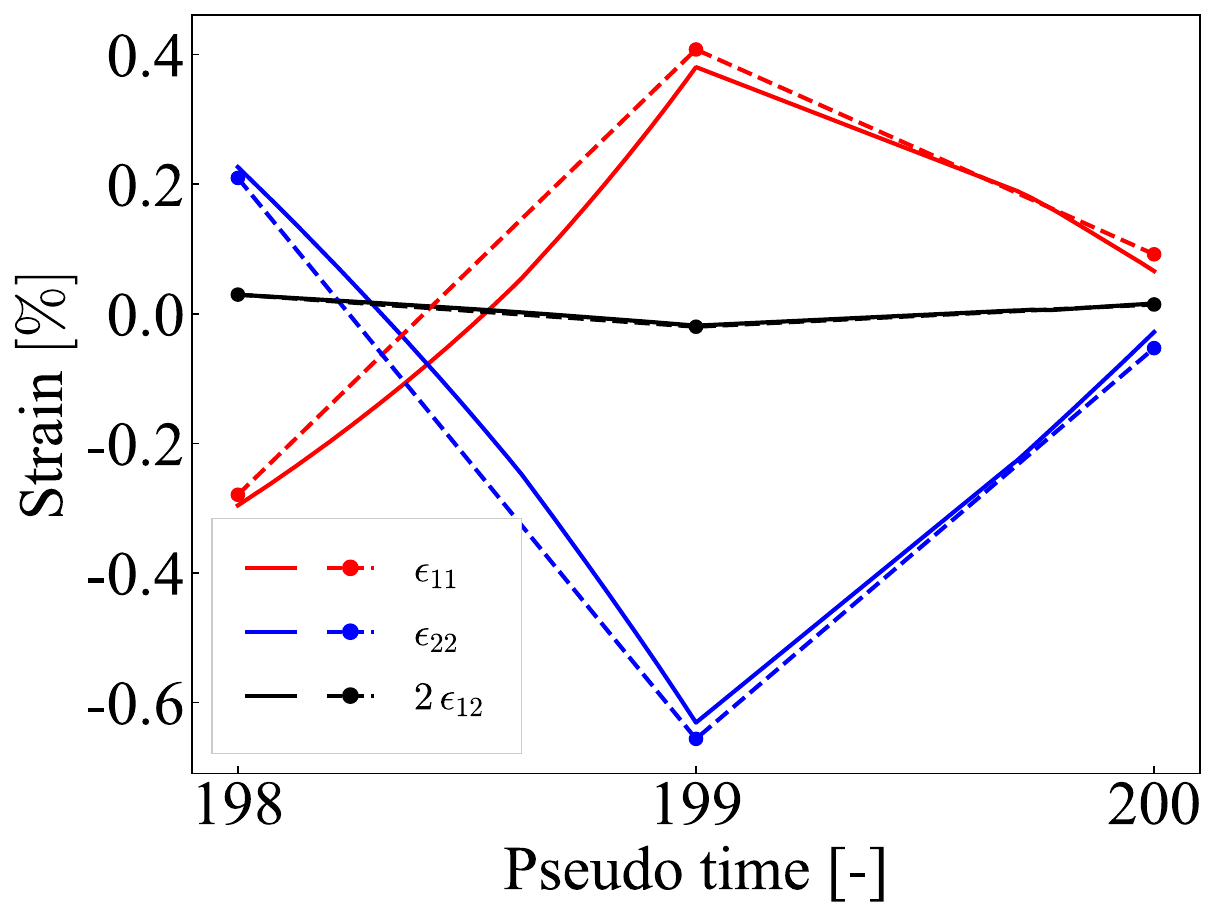}
        \caption{}
        \label{fig:strain_histories_el_79_incr_halfcyle_1}
    \end{subfigure}
    \hspace*{0.01\textwidth} 
    \begin{subfigure}{0.48\textwidth}
        \includegraphics[width=1\textwidth]{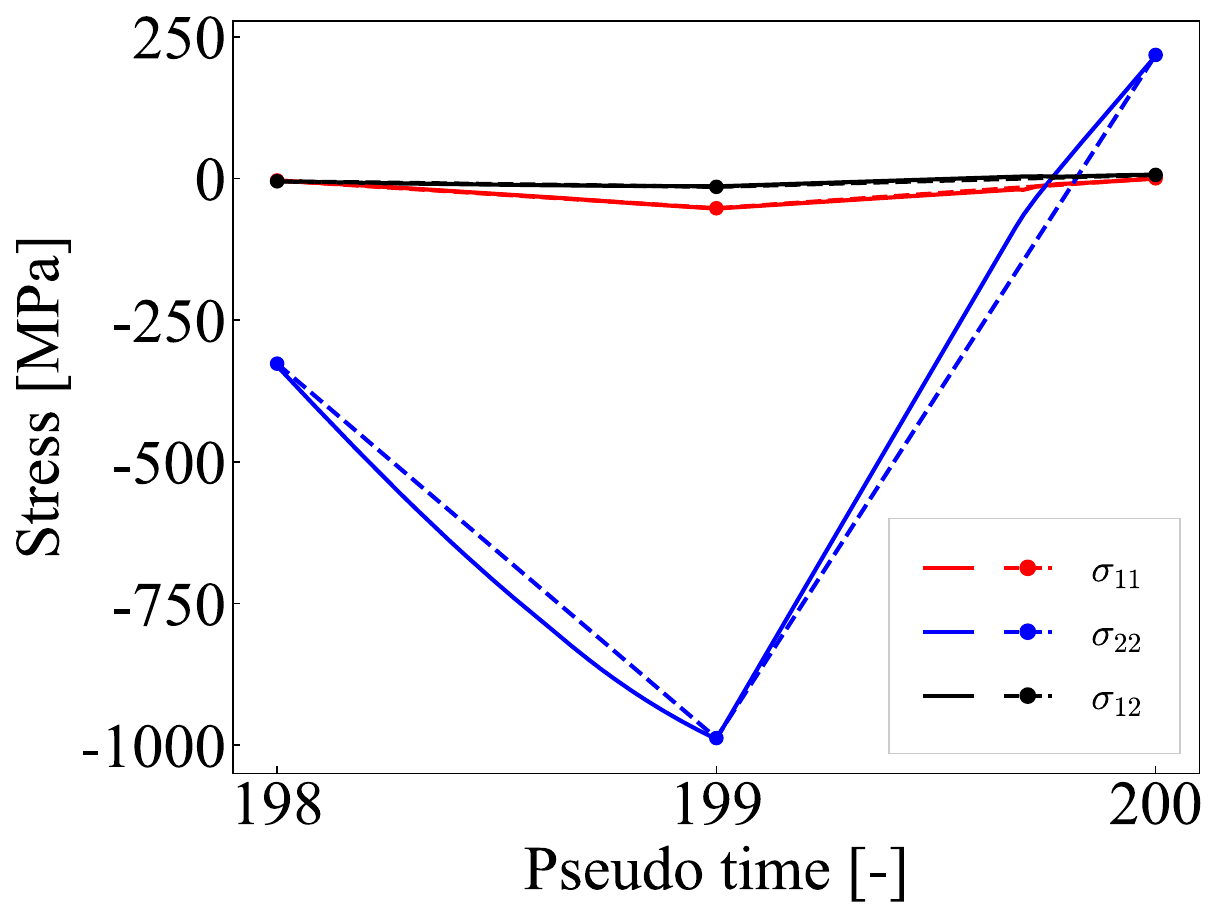}
	\caption{}
	\label{fig:stress_histories_el_79_incr_halfcyle_1}
    \end{subfigure} 
\caption{(a) Strain histories and (b) stress histories for the critical element (indicated in Figure~\ref{fig:FE_setup}) in the last cycle. The solid lines correspond to the reference FE model, and the dashed lines refer to the NN-based FE model with 1 displacement increment per half-cycle.}
\label{fig:stress_strain_history_el79}
\end{figure*}

The von Mises stress, $\sigma^{\rm{vM}}$, distributions in cycles 50 and 100 are illustrated in Figures~\ref{fig:vonMises_stress_ref_ML_combined}a and~\ref{fig:vonMises_stress_ref_ML_combined}c at the peak displacements for the reference and NN-based FE models. Overall, the results from the latter show a close agreement with the reference FE model results in both cycles. To further assess the differences between the models, spatial distributions of the absolute errors in $\sigma^{\rm{vM}}$ are presented in Figures~\ref{fig:vonMises_stress_ref_ML_combined}b and~\ref{fig:vonMises_stress_ref_ML_combined}d. In cycle 50, the maximum absolute errors are 8.5 MPa and 3.6 MPa for the NN-based model using 5 and 1 increments per half-cycle, respectively. The corresponding errors in cycle 100 are 4.4 MPa and 8.0 MPa. Relative to the maximum $\sigma^{\rm{vM}}$ values in the plate, approximately 732 MPa in cycle 50 and 821 MPa in cycle 100, these errors show that the FE models employing the NN-based integration with both increment numbers provide accurate predictions. To showcase the predictions of the NN-based model over the 100 simulated cycles in the highly stressed region near the hole, Figure~\ref{fig:FE_vonMises_el79} compares $\sigma^{\rm{vM}}$ values from the reference and NN-based FE models at the peak displacements for the critical element indicated in Figure~\ref{fig:FE_setup}. Despite the maximum absolute errors of 3.6 MPa and 3.5 MPa for 5 and 1 displacement increments per half-cycle, the predictions of the explicit NN-based integration are very promising. Further, the strain histories during the last cycle for the critical element are shown in Figure~\ref{fig:strain_histories_el_79_incr_halfcyle_1}, comparing the reference model and the NN-based FE model using 1 increment per half-cycle. Considering the reference results, the strain paths are nonlinear during the displacement increments, which violates our assumption of a linear variation of $\ts\eps$ within a time step. Nevertheless, as shown in Figure~\ref{fig:stress_histories_el_79_incr_halfcyle_1}, the predicted stresses obtained with the NN-based model are in close agreement with those from the reference model. It should be noted that $\sigma_{11}$ and $\sigma_{12}$ tend toward zero as the mesh is further refined near the hole. 

\begin{figure}[!t]
    \centering
    \includegraphics[width=0.95\textwidth]{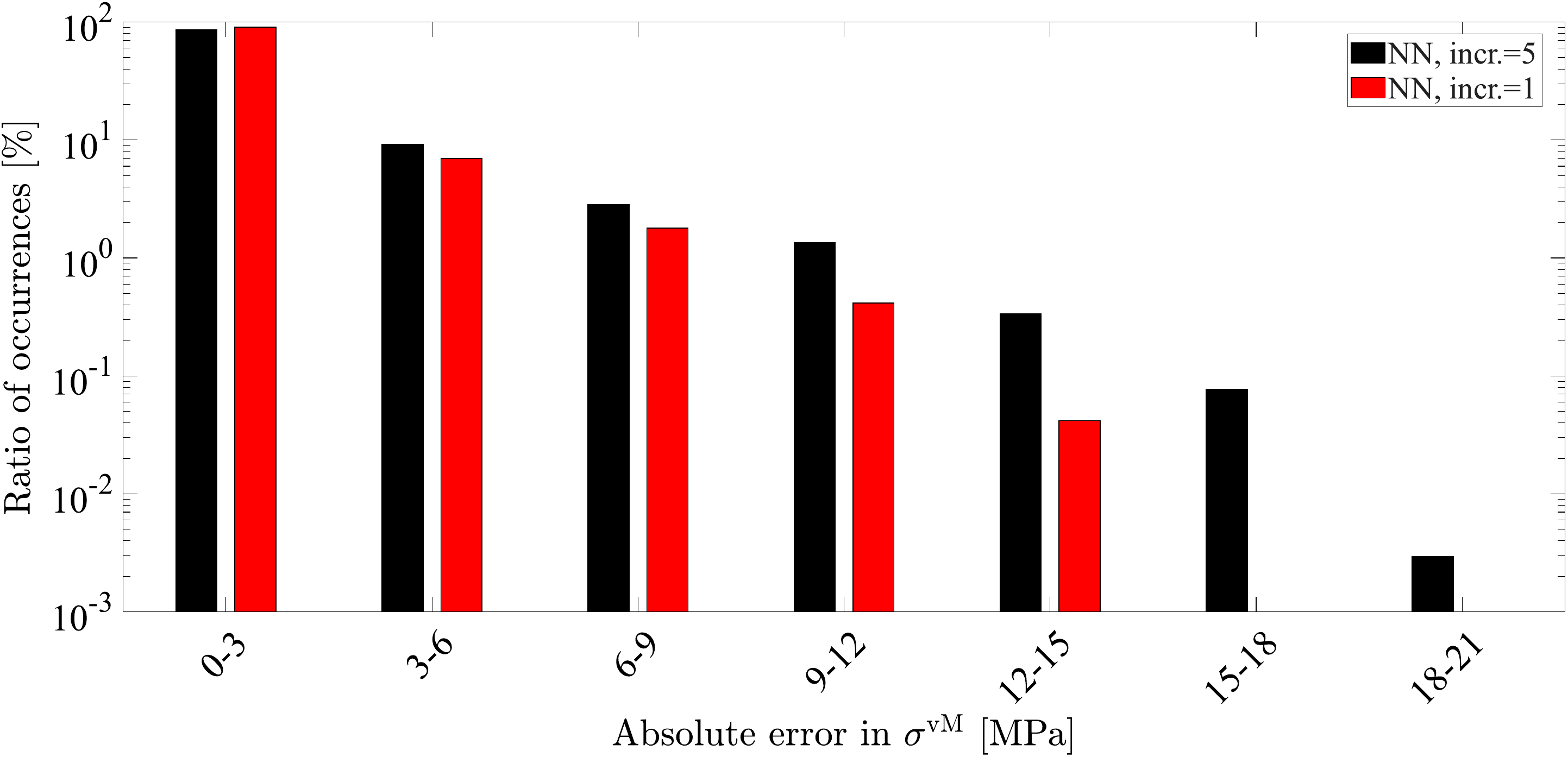}
    \caption{Distribution of absolute errors in von Mises stresses ($\sig^{\rm{vM}}$) at the peak displacements over 100 cycles, considering all finite elements. NN, incr.=5 and NN, incr.=1 refer to the FE models using the NN-based integration scheme with 5 and 1 displacement increments per half-cycle, respectively.}
    \label{fig:histogram_von_mises_stress_error}
\end{figure}

Figure~\ref{fig:histogram_von_mises_stress_error} presents the distribution of the absolute errors in $\sigma^{\rm{vM}}$ over all cycles and finite elements. 
For both numbers of displacement increments, most errors are concentrated in the lowest interval of $0-3$ MPa, accounting for approximately $85\%$ to $90 \%$ of all occurrences. The relative frequency decreases rapidly as the error increases, with only a small fraction of cases showing errors larger than 9 MPa. The maximum absolute errors considering 5 and 1 displacement increments are 18.1 MPa and 14.3 MPa, respectively. Compared to the maximum $\sigma^{\rm{vM}}$ of 885.3 MPa in the reference model, these errors indicate accurate predictions of the NN-based model.

Finally, the computational efficiency of the FE simulations using the reference and NN-based models has been compared in terms of Central Processing Unit (CPU) time. Each FE simulation was conducted 5 times, and the averaged results are presented in Table~\ref{tab:cpu_time}. All simulations were performed on a Windows laptop equipped with an Intel Core Ultra 7 165H processor and 32 GB of RAM.
The FE simulations employing the NN-based integration with 5 and 1 displacement increments per half-cycle are approximately 14.6 and 34.6 times faster than the reference model, respectively. Further, we have reduced the number of displacement increments for the reference model, using the backward Euler time integration scheme. This has been done until the maximum absolute error in $\sigma^{\rm{vM}}$, evaluated over all cycles and finite elements relative to the highly resolved solution using 100 displacement increments per half-cycle, became comparable to that obtained with the FE models using the NN-based integration.
For this purpose, the number of displacement increments for the reference model has been decreased to 8 per half-cycle. This has resulted in the maximum absolute error of 16.1 MPa in $\sigma^{\rm{vM}}$. It should be mentioned that lowering the number of increments to 5 did not give convergence during the displacement iterations. When compared with the reference model using 8 increments per half-cycle, the NN-based simulations with 5 and 1 increments are roughly 1.4 and 3.3 times faster, respectively. 

\begin{table}[h!]
    \caption{CPU time comparison between the reference and NN-based FE models.}
    \centering \footnotesize
    \setlength{\tabcolsep}{8pt}
    \begin{threeparttable}
    \begin{tabular}{l c c c c} 
    \hline
    Model & \multicolumn{2}{c}{Reference FE model} & \multicolumn{2}{c}{NN-based FE model} \\
    \hline
    No. displacement increments & 100 & 8 & 5 & 1 \\
    \hline
    Averaged CPU time [sec] & 672\textsuperscript{*} & 63.8 & 46.0 & 19.4 \\
    \hline
    Max error in $\sigma^{\rm{vM}}$ [MPa] & - & 16.1 & 18.1 & 14.3 \\
    \hline
    \end{tabular}
    \begin{tablenotes}
        \footnotesize
        \item[*] Using the Abaqus built-in material model has resulted in an averaged CPU time of 670 [sec].
    \end{tablenotes}
    \end{threeparttable}
    \label{tab:cpu_time}
\end{table}

\section{Concluding remarks}
\label{sec:summary}

In this contribution, we have presented an NN-based framework to accelerate the time integration of DAEs in rate-independent nonlinear constitutive models. 
The key features of the framework are
\begin{itemize}
    \item Exact encoding of the separation between elastic and plastic responses.
    \item State variable updates formulated by neural networks with invariants as inputs.
    \item Pre-identification of sufficient and necessary evolution directions based on the training data.
    \item Exact fulfillment of the plastic consistency condition.
\end{itemize}

We have chosen a prototype material model with the von Mises yield function and nonlinear kinematic hardening. Under pulsating proportional multiaxial loading, we have generated training data for the FFNNs using the prototype model, integrated with the implicit backward Euler time integration algorithm. 
While only 10 loading cycles were used to generate training data, the proposed NN-based time integration scheme gave very good accuracy for material point simulations. 
Even when taking a single strain increment per half-cycle for 100 loading cycles, it achieved good predictions.
Furthermore, the framework's performance has been evaluated in cyclic finite element (FE) simulations. The von Mises stress predictions were accurate, with a maximum error of less than $2\%$ of the maximum stress over 100 cycles when using a single displacement increment per half-cycle. Moreover, the NN-based time integration is roughly 3 times faster than the reference FE model when taking a single displacement increment per half-cycle. 

\section{Acknowledgements}
This work is part of the ongoing activities within the National Center of Excellence CHARMEC (\url{www.chalmers.se/charmec}). Parts of the study have been funded by Europe’s Rail project IAM4RAIL under grant agreement No. 101101966. Parts of the computations were enabled by resources provided by Chalmers e-Commons.

\clearpage
\appendix
\section*{Appendix}

\section{Algorithmic tangent stiffness}
\label{appendix:ats_tensor}

This section presents the derived analytical expression for the Algorithmic Tangent Stiffness (ATS). The ATS tensor is defined as
\begin{equation}
    \tf{E}^{\mathrm{a}} = \diff[\sig]{\eps} = \diff[\sig]{\Delta\eps}
\end{equation}
Considering that $\sig = \oldtime{\sig} + \tf{E}^{\rm{e}} : \left[\Delta\eps - \Delta\eps\pl\right]$, the ATS tensor is expressed as
\begin{equation}
    \tf{E}^{\mathrm{a}} = \tf{E}^{\mathrm{e}} - \tf{E}^{\mathrm{e}} : \diff[\Delta\eps\pl]{\Delta\eps} = \tf{E}^{\mathrm{e}} - \tf{E}^{\mathrm{e}} : \pdiff[\Delta\eps\pl] {\sig\trial} : \tf{E}^{\mathrm{e}} 
\end{equation}
With Equation~\ref{eq:statevar_incr_final}, the differentiation of $\Delta\eps\pl$ with respect to $\sig\trial$ leads to
\begin{equation}
    \pdiff[\Delta\eps\pl] {\sig\trial} = c \, \diff[{\tn{g}}_{\epsilon^{\rm p}}\supscr{NN}]{\sig\trial} + {\tn{g}}_{\epsilon^{\rm p}}\supscr{NN} \otimes \diff[c]{\sig\trial}
\end{equation}
$\rm{d}{\tn{g}}_{\epsilon^{\rm p}}\supscr{NN}/\rm{d}{\sig\trial}$ is derived as
\begin{align}
    \diff[{\tn{g}}_{\epsilon^{\rm p}}\supscr{NN}]{\sig\trial} &=
    \mathtt{SF}_{\epsilon^{\rm{p}}} \Biggl[\neuralnet_{\, \eps\pl, \, 1}\left(\set{\hat{J}}\right) \frac{1}{{\norm{\sig\trialdev}}_{\rm{max}}}\, \tf{I}^{\dev} \,
    \nonumber  +  \neuralnet_{\, \eps\pl, \, 2}\left(\set{\hat{J}}\right) \frac{1}{{\norm{\sig^{\rm{lim, \, dev}}}}_{\rm{max}}} \, \left[\pdiff[\sig^{\rm{lim, \, dev}}]{\sig\trialdev} : \tf{I}^{\dev}\right] + \nonumber \\ 
    &\hat{\sig}\trialdev \otimes {\color{red}\biggl[}
    \pdiff[\neuralnet_{\, \eps\pl, \, 1}\left(\set{\hat{J}}\right)]{J_{\rm{tr, \, tr}}} \, 2 \, \sig\trialdev + \pdiff[\neuralnet_{\, \eps\pl, \, 1}\left(\set{\hat{J}}\right)]{J_{\rm{tr, \, lim}}} \, \Bigl[\sig^{\rm{lim, \, dev}} + \sig\trialdev : \pdiff[\sig^{\rm{lim, \, dev}}]{\sig\trialdev} \Bigr] + \pdiff[\neuralnet_{\, \eps\pl, \, 1}\left(\set{\hat{J}}\right)]{J_{\rm{tr, \, \beta}}} \, \oldtime{\backstress}
    {\color{red}\biggr]} + \nonumber \\
    & \hat{\sig}^{\rm{lim, \, dev}} \otimes {\color{blue}\biggl[}
    \pdiff[\neuralnet_{\, \eps\pl, \, 2}\left(\set{\hat{J}}\right)]{J_{\rm{tr, \, tr}}} \, 2 \, \sig\trialdev +  
    \pdiff[\neuralnet_{\, \eps\pl, \, 2}\left(\set{\hat{J}}\right)]{J_{\rm{tr, \, lim}}} \, \Bigl[\sig^{\rm{lim, \, dev}} + \sig\trialdev : \pdiff[\sig^{\rm{lim, \, dev}}]{\sig\trialdev} \Bigr] +
    \pdiff[\neuralnet_{\, \eps\pl, \, 2}\left(\set{\hat{J}}\right)]{J_{\rm{tr, \, \beta}}} \, \oldtime{\backstress}
    {\color{blue}\biggr]}  + \nonumber \\
    &\oldtime{\hat{\backstress}} \otimes {\color{green}\biggl[}
    \pdiff[\neuralnet_{\, \eps\pl, \, 3}\left(\set{\hat{J}}\right)]{J_{\rm{tr, \, tr}}} \, 2 \, \sig\trialdev +  
    \pdiff[\neuralnet_{\, \eps\pl, \, 3}\left(\set{\hat{J}}\right)]{J_{\rm{tr, \, lim}}} \, \Bigl[\sig^{\rm{lim, \, dev}} + \sig\trialdev : \pdiff[\sig^{\rm{lim, \, dev}}]{\sig\trialdev} \Bigr] +
    \pdiff[\neuralnet_{\, \eps\pl, \, 3}\left(\set{\hat{J}}\right)]{J_{\rm{tr, \, \beta}}} \, \oldtime{\backstress}
    {\color{green}\biggr]}
    \Biggr] \label{eq:dgepsp_dsig}
\end{align}
where $\partial{\sig^{\rm{lim, \, dev}}}/\partial{{\sig\trialdev}}$ is expressed as
\begin{align}
    \pdiff[\sig^{\rm{lim, \, dev}}]{\sig\trialdev} = 
    \pdiff{\sig\trialdev} \biggl[\oldtime{\sig\dev} + s \left[\sig\trialdev -\oldtime{\sig\dev}\right]\biggr]= s \, \tf{I} + \left[\sig\trialdev - \oldtime{\sig\dev}\right] \otimes \diff[s]{\sig\trialdev}
\end{align}
Considering Equation~\ref{eq:s_param}, $\dif s / \dif \sig\trialdev$ is derived from 
\begin{align}
    &\diff[f_s]{\sig\trialdev} = \left.\pdiff[f_s]{\sig\trialdev}\right\vert_{s} + \left.\pdiff[f_s]{s}\right\vert_{\sig\trialdev} \diff[s]{\sig\trialdev} = 0 \rightarrow \\
    &\diff[s]{\sig\trialdev} = - \left[\pdiff[f_s]{s}\right]^{-1} \, \pdiff[f_s]{\sig\trialdev}
\end{align}
where
\begin{align}
    &\pdiff[f_s]{s} = 3 \, \Biggl[ \left[ \oldtime{\sig\dev} - \oldtime{\backstress} \right] : \left[\sig\trialdev - \oldtime{\sig\dev} \right] + s \, \left[\sig\trialdev - \oldtime{\sig\dev} \right] : \left[\sig\trialdev - \oldtime{\sig\dev} \right] 
    \Biggr]
\end{align}
and
\begin{align}
    \pdiff[f_s]{\sig\trialdev} = 3 \, s \, \Biggl[ \oldtime{\sig\dev + s \, \left[\sig\trialdev - \oldtime{\sig\dev}\right] -\oldtime{\backstress}}
    \Biggr]
\end{align}
Considering Equation~\ref{eq:c_factor}$, \rm{d} c / \rm{d}{\sig\trial}$ is derived from
\begin{align}
    &\diff[f_c]{\sig\trial} = \left.\pdiff[f_c]{\sig\trial}\right\vert_{c} + \left.\pdiff[f_c]{c}\right\vert_{\sig\trial} \diff[c]{\sig\trial} = 0 \rightarrow \\
    &\diff[c]{\sig\trial} = - \left[\pdiff[f_c]{c}\right]^{-1} \, \pdiff[f_c]{\sig\trial}
\end{align}
where 
\begin{align}
&\pdiff[f_c]{c} = -3 \, \Biggl[ \left[\sig\trialdev - \oldtime{\backstress}\right]  : \left[2 \, G \, {\tn{g}}_{\epsilon^{\rm p}}\supscr{NN} -2/3 \, H_{\rm{kin}} \, {\tn{g}}_{b}\supscr{NN} \right] + c \, \norm{2 \, G \, {\tn{g}}_{\epsilon^{\rm p}}\supscr{NN} -2/3 \, H_{\rm{kin}} \, {\tn{g}}_{b}\supscr{NN}}^{2}
\Biggr]
\end{align}
and 
\begin{align}
\pdiff[f_c]{\sig\trial} = 3 \, \Biggl[\tf{I}^{\rm{dev}} - c \, \biggl[ 2 \, G \, \pdiff[{\tn{g}}_{\epsilon^{\rm p}}\supscr{NN}]{\sig\trial} -2/3 H_{\rm{kin}} \,  
\pdiff[{\tn{g}}_{b}\supscr{NN}]{\sig\trial}\biggr]\, \Biggr] : 
\Biggl[\sig\trialdev - \oldtime{\backstress} - c \, \biggl[2 \, G {\tn{g}}_{\epsilon^{\rm p}}\supscr{NN} - 2/3 \, H_{\rm{kin}} \, {\tn{g}}_{b}\supscr{NN}\biggr] 
\Biggr]
\end{align}
$\rm{d}{\tn{g}}_{b}\supscr{NN}/\rm{d}{\sig\trial}$ is derived in a similar way as that for $\rm{d}{\tn{g}}_{\epsilon^{\rm p}}\supscr{NN}/\rm{d}{\sig\trial}$. It should be noted that the required derivatives of the outputs of the FFNNs, i.e., $\neuralnet_{\, \statevar_i, \, 1}$, $\neuralnet_{\, \statevar_i, \, 2}$, and $\neuralnet_{\, \statevar_i, \, 3}$, with respect to the inputs have been calculated analytically using chain rule. Further details can be found in, e.g.,~\cite{weber2023}.

\section{Additional results}
\label{appendix:additional_results}

\begin{figure}[!h]
    \centering
    \includegraphics[width=0.88\textwidth]{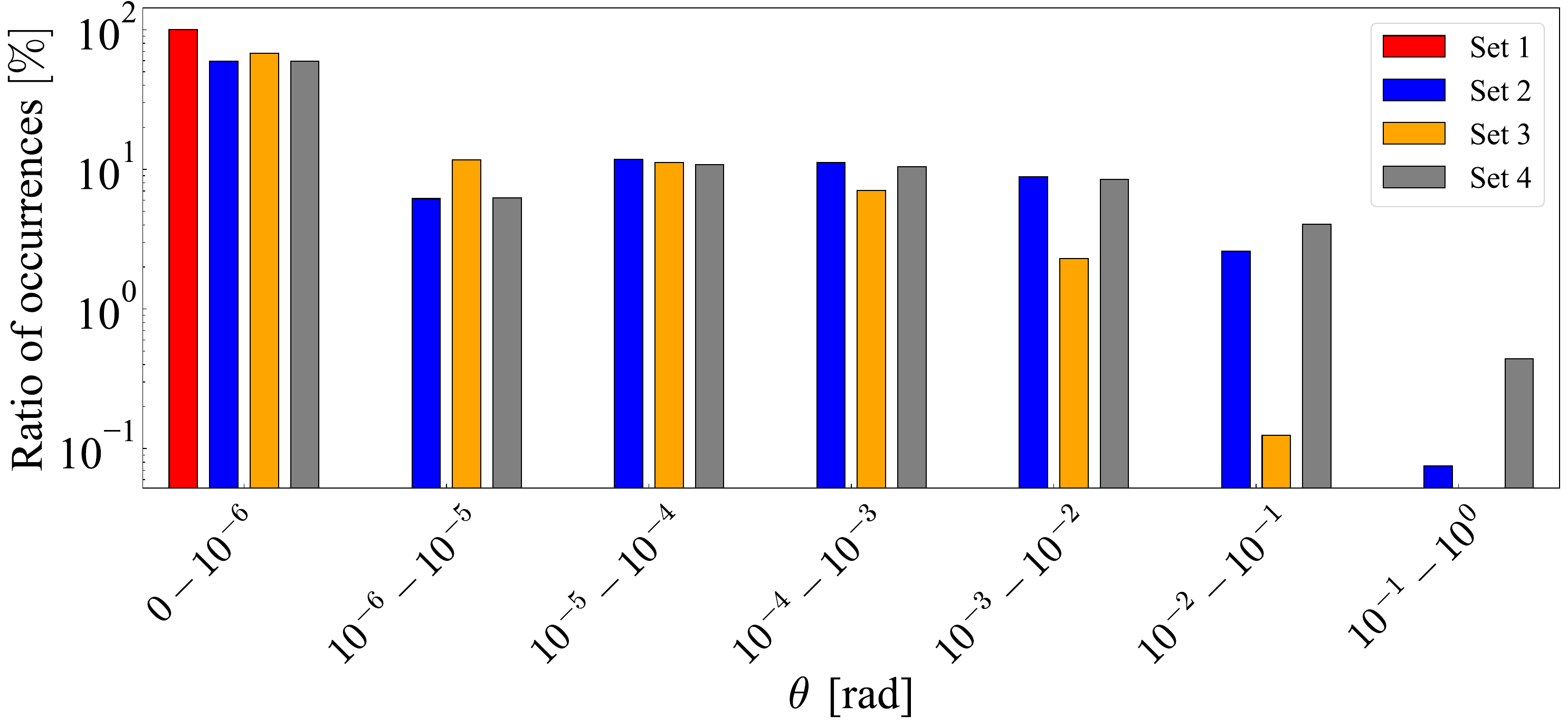}
    \caption{The distribution of the computed angles, $\theta$, between the target and predicted $\Delta\ts{b}$ considering all training data for four sets of evolution directions: set 1: $\left\{{\sig}\trialdev, {\sig}^{\rm{lim, \ dev}},\oldtime{{\backstress}}\right\}$, set 2: $\left\{{\sig}\trialdev, {\sig}^{\rm{lim, \, dev}}\right\}$, set 3: $\left\{{\sig}\trialdev, \oldtime{{\backstress}}\right\}$, and set 4: $\left\{{\sig}^{\rm{lim, \, dev}}, \oldtime{{\backstress}}\right\}$.}
    \label{fig:angles_beta}
\end{figure}

This section provides supplementary results for Sections~\ref{subsec:select_evolution_directions} and~\ref{subsec:material_point_sim_res}. 
Figure~\ref{fig:angles_beta} shows the computed $\theta_i$ values versus the relative number of occurrences considering $\Delta\ts{b}$, using the proposed method described in Section~\ref{subsec:select_evolution_directions}.

\begin{figure}[!t]
\vspace*{-12cm}
    \centering
    \begin{subfigure}{0.47\textwidth}
        \includegraphics[width=1\textwidth]{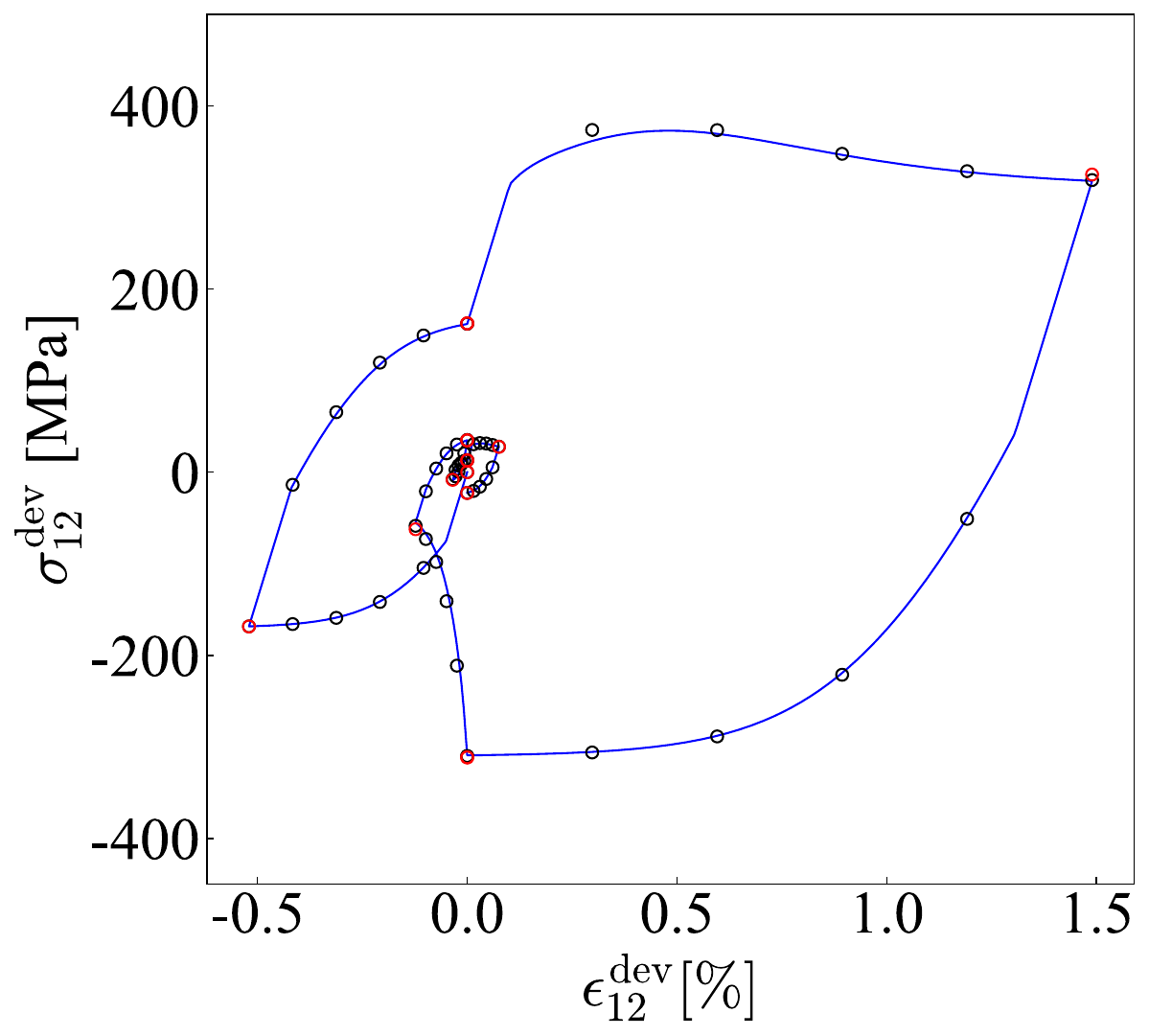}
        \caption{Training data}
        \label{fig:shear_response_material_point_sim_train}
	\end{subfigure}
    \hspace*{0.03\textwidth} 
    \begin{subfigure}{0.47\textwidth}
        \includegraphics[width=1\textwidth]{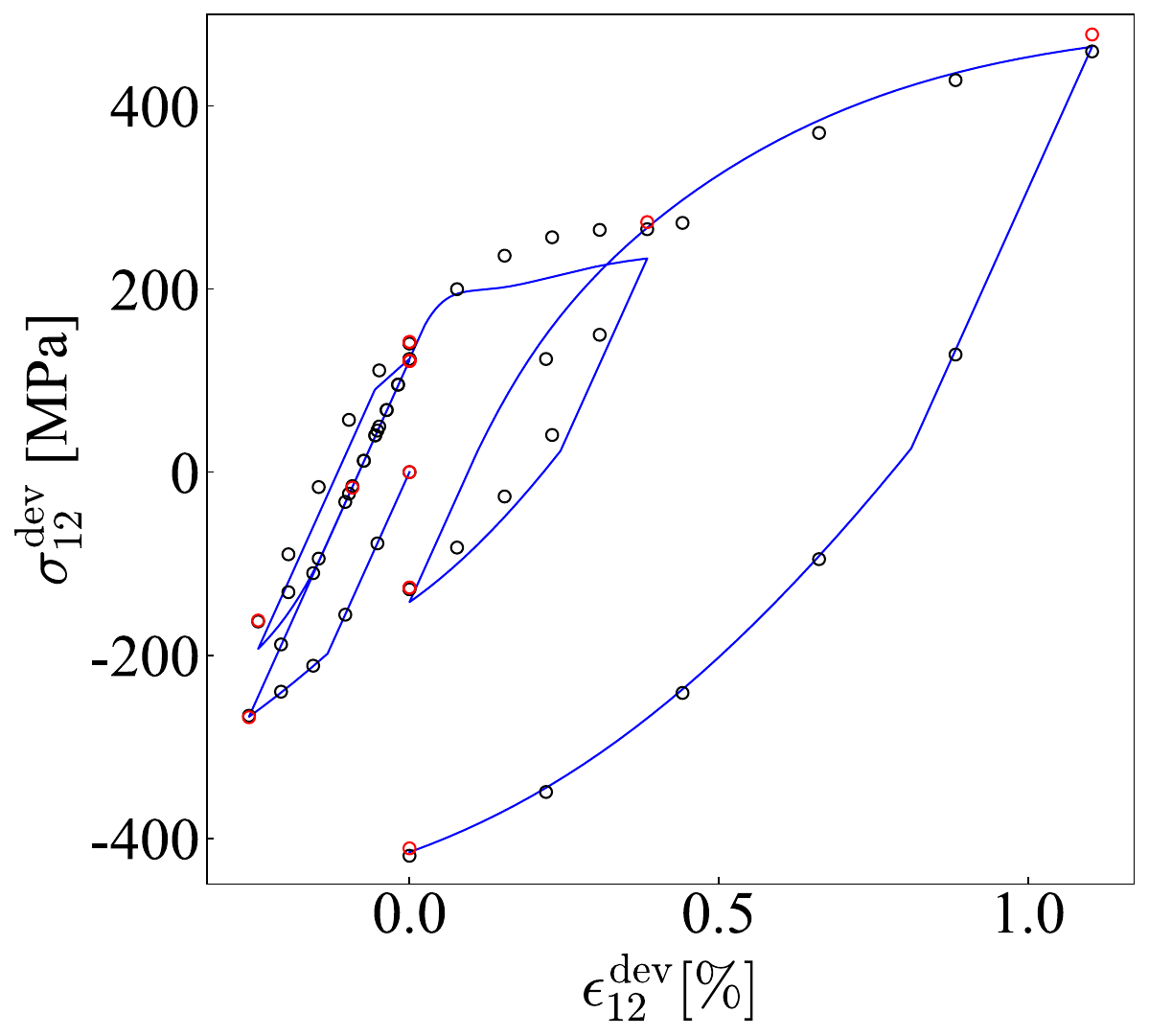}
        \caption{Test data}
        \label{fig:shear_response_material_point_sim_test}
	\end{subfigure}  
\caption{The shear stress-strain responses from the reference material model (blue) and the NN-based model. The black and red circles correspond to the results from the NN-based model with 5 and 1 strain increments per half-cycle, respectively. Only the first 5 cycles are shown.}
\label{fig:shear_response_material_point_sim}
\end{figure}

Figure~\ref{fig:shear_response_material_point_sim_train} shows the deviatoric shear responses over the first 5 cycles for the reference and NN-based models, considering the training data. The corresponding responses for the test data are presented in Figure~\ref{fig:shear_response_material_point_sim_test}.

\clearpage
\bibliographystyle{elsarticle-num}
\setlength{\parskip}{0em}
\bibliography{references}
\end{document}